\documentclass[structabstract]{aa}  
\usepackage{graphicx}
\usepackage{txfonts}
\usepackage{natbib}
\usepackage{lscape}
\begin{document}
   \title{Herschel/HIFI observations of O-rich AGB stars : molecular
inventory\thanks{Herschel is an ESA space observatory with science instruments 
provided by European-led Principal Investigator consortia and with 
important participation from NASA.}
}
\author{K. Justtanont\inst{1}
\and T. Khouri\inst{2}
\and M. Maercker\inst{3,16}
\and J. Alcolea \inst{4}
\and L. Decin\inst{5,2}
\and H. Olofsson \inst{1}
\and F.~L. Sch\"oier \inst{1}\thanks{deceased}
\and V. Bujarrabal \inst{6}
\and A.P. Marston \inst{7}
\and D. Teyssier \inst{7}
\and J. Cernicharo \inst{8}
\and C. Dominik \inst{2,9}
\and A. de Koter \inst{2,10}
\and G. Melnick \inst{11}
\and K. M. Menten \inst{12}
\and D. Neufeld \inst{13}
\and P. Planesas \inst{4,14}
\and M. Schmidt \inst{15}
\and R. Szczerba \inst{15}
\and R. Waters \inst{2,5}
}
\institute{Onsala Space Observatory, Dept. Earth and Space Science,
Chalmers University of Technology, S-439 92 Onsala, Sweden\\
 \email{kay.justtanont@chalmers.se}
\and Sterrenkundig Instituut Anton Pannekoek, University of Amsterdam,
Science Park 904, NL-1098 Amsterdam, The Netherlands
%\and Department of Astronomy, AlbaNova University Center, Stockholm
%University, SE--10691 Stockholm, Sweden
\and University of Bonn, Argelander-Institut f\"{u}r Astronomie, 
Auf dem H\"{u}gel 71, D-53121 Bonn, Germany
\and Observatorio Astron\'omico Nacional (IGN), Alfonso XII N$^{\circ}$3,
E-28014 Madrid, Spain
\and Instituut voor Sterrenkunde, Katholieke Universiteit Leuven, 
Celestijnenlaan 200D, 3001 Leuven, Belgium
\and Observatorio Astron\'omico Nacional. Ap 112, E-28803
Alcal\'a de Henares, Spain
\and European Space Astronomy Centre, ESA, P.O. Box 78, E-28691
Villanueva de la Ca\~nada, Madrid, Spain
\and CAB, INTA-CSIC, Ctra de Torrej\'on a Ajalvir, km 4,
28850 Torrej\'on de Ardoz, Madrid, Spain
\and Department of Astrophysics/IMAPP, Radboud University Nijmegen,   
Nijmegen, The Netherlands
\and
Astronomical Institute, Utrecht University,
Princetonplein 5, 3584 CC Utrecht, The Netherlands
\and Harvard-Smithsonian Center for Astrophysics, Cambridge, MA 02138, USA
\and Max-Planck-Institut f{\"u}r Radioastronomie, Auf dem H{\"u}gel 69,
D-53121 Bonn, Germany
\and Johns Hopkins University, Baltimore, MD 21218, USA
\and Joint ALMA Observatory, Alonso de Córdova 3107, Vitacura, Santiago, Chile
\and N. Copernicus Astronomical Center, Rabia{\'n}ska 8, 87-100 Toru{\'n}, 
Poland
\and European Southern Observatory, Karl Schwarzschild Str. 2, Garching bei
M\"{u}nchen, Germany
}
   \date{Received 20 June 2011 ; accepted 14 Nov 2011 }
\abstract
% context : optional
{}
%aims : mandatory
{%Herschel-HIFI spectra 
Spectra, taken with the heterodyne instrument, HIFI, aboard the
Herschel Space Observatory, 
of O-rich asymptotic giant branch (AGB) stars which form part of the
guaranteed time key program HIFISTARS are presented.
The aim of this program is to study the dynamical structure, 
mass-loss driving mechanism, and chemistry of the outflows from
AGB stars as a function of chemical composition and initial mass.
}
%methods : mandatory
{We used the HIFI instrument to observe nine AGB stars, mainly 
in the H$_{2}$O and high rotational CO lines.% and present integrated
%line intensities, line luminosities and line widths.
We investigate the correlation between line luminosity, line
ratio and mass-loss rate, line width and excitation energy.
}
%results : mandatory
{
A total of nine different molecules, along with some of their 
isotopologues have been identified, covering a
wide range of excitation temperature. Maser emission is detected
in both the ortho- and para-H$_{2}$O molecules.
The line luminosities of ground state lines of ortho- and para-H$_{2}$O,
the high-J CO and NH$_{3}$ lines show a clear correlation with mass-loss 
rate. The line ratios of H$_{2}$O and NH$_{3}$ relative to CO J=6-5 
correlate with the mass-loss rate while ratios of higher CO lines 
to the 6-5 is independent of it.
In most cases, the expansion velocity derived 
from the observed line width of highly excited transitions formed
relatively close to the stellar photosphere is lower than that of lower
excitation transitions, formed farther out, pointing to an accelerated
outflow. In some objects, the vibrationally excited H$_{2}$O and SiO
which probe the acceleration zone suggests the wind reaches its 
terminal velocity already in the innermost part of the envelope, i.e.,
the acceleration is rapid. 
Interestingly, for R Dor we find indications of a deceleration
of the outflow in the region where the material has already escaped
from the star.
}
% conclusion : optional
{}

\maketitle

\section{Introduction}

Asymptotic giant branch (AGB) stars represent a late stage of
stellar evolution when the nuclear burning in the core has ceased.
The dominant factor governing the rest of their evolution is the
mass loss from the surface \citep{habing96}. 
%****
%Because of pulsation, the photosphere becomes extended and 
%the density is high enough for dust grains to condense out. 
%****
Presently, the initial 
mechanism(s) driving the wind which leads to dust formation
in O-rich stars is not fully
understood.
Model calculations of dynamics of the photosphere of AGB stars
show that shockwaves arising from pulsation can levitate the
material and produce an extended atmosphere \citep[e.g.,][]
{bowen88, hoefner92, hoefner97}. The latter authors showed
that the models lead to dust formation in the atmosphere.
Although the efficiency of dust formation
in the outflow of AGB stars has been questioned \citep[see e.g.,][]
{woitke06}, \citet{hoefner08} proposed the wind is driven by 
micron-size grains which are Fe-poor. \citet{mattsson08} also show that
for AGB stars with a low metallicity, pulsation can lead to an intense
mass-loss rate.%}
%****
Due to the larger cross sections of dust grains compared to molecules,
dust efficiently absorbs stellar radiation and is accelerated outwards,
dragging the gas along with it \citep{gold76,just94,decin06,ramstedt08}. 
This effectively establishes a circumstellar envelope around the star.

The circumstellar envelope of an AGB star is an active site for chemistry.
The slowly expanding wind ($\rm v_{e} \sim$ 10-15 km s$^{-1}$) produced by
a constant, isotropic mass-loss rate, 
%a circumstellar envelope around AGB stars 
is an ideal laboratory to study physical and chemical processes during 
stellar evolution.
In thermal equilibrium, the relative abundance of carbon and
oxygen determines the composition of the molecules and dust species formed.
For a carbon-rich star with 
C/O $>$ 1, carbonaceous molecules form as well as 
amorphous carbon and SiC dust after all the oxygen is locked up in 
the most abundant trace molecule CO. 
In contrast, for a star with  C/O $<$ 1, i.e., O-rich (M-type)
AGB stars, H$_{2}$O and CO are the main 
gas components,
along with silicate dust. However, close to the photosphere,
shock waves due to stellar pulsations and/or the stellar radiation field
can induce non-thermal equilibrium, rendering this simple 
picture more complicated \citep{cherchneff06}.
As an example, the canonical C-rich AGB star, IRC+10216 has been
found to exhibit H$_{2}$O emission \citep{melnick01, hase06,
decin10nat}. Data obtained with Herschel-HIFI have recently shown that
H$_{2}$O is quite prevalent in
C-stars \citep{neufeld11}.

HIFISTARS is the guaranteed time key program aimed at 
studying circumstellar envelopes around evolved stars using
the Heterodyne Instrument \citep[HIFI,][]{degraauw10} aboard
the Herschel Space Observatory \citep{pilbratt10}. The program
aims to study stars with a large range of mass-loss rates
($10^{-7}-10^{-4}$ $M_{\odot}$ yr$^{-1}$), differing chemistry
(C/O$<$1, C/O$\sim$1, C/O$>$1) and initial mass (low- and intermediate mass
stars and supergiants), as well as different evolutionary 
(AGB to planetary nebula) phases. In this paper, we report all the
observations done on O-rich AGB stars in our program. 
Observations and data reduction
are presented in Sect.~2 and the lines detected
and the first-cut interpretation are discussed in Sect.~3.
The results are summarised in Sect.~4.

\section{Observations}

Our HIFI observations were
carried out using dual-beam switch mode with a throw of 3$^{\prime}$
and a slow chopping. 
The full bandwidth of HIFI of 4\,GHz was utilized
using the wide-beam-spectrometer backend. A total of 16
different frequency settings have been chosen to cover a number
of expected strong lines of H$_{2}$O and CO for the purpose
of sampling different regions of the warm  
circumstellar envelope in objects. 
The data were calibrated
using the standard Herschel pipeline, HIPE, and reprocessed for those
which had been processed with the pipeline version earlier than 4.0.
Data were taken using two orthogonal polarizations: horizontal and vertical.
%polarizations. 
A resulting spectrum is an average of these two polarizations
which is then rebinned to a 1 km s$^{-1}$ resolution 
(Figs.~\ref{fig_spec}-~\ref{fig_rcas}).
However, a number of spectra in the THz-band (HIFI bands 6 and 7) were 
affected by the ripples, especially in the v-polarization. 
These are thought to be due to standing waves in the hot electron
bolometer (HEB) mixers. In these cases, 
we rebinned only the h-polarization spectrum. The spectra have been
corrected for the beam efficiency, $\eta_{\rm mb}$
\begin{equation}
 \eta_{\rm mb} = \eta_{\rm mb,\rm 0} \exp (-(4\pi\sigma/\lambda)^2)
\end{equation}
where $\sigma$ is the surface accuracy (= 3.8\,$\mu$m), 
$\lambda$ is the wavelength of the transition and
$ \eta_{\rm mb,\rm 0}$ is a correction factor of 0.76, except for
the frequency range of 1120-1280\,GHz (HIFI band 5) where this value is 0.66
(Olberg, 2010, http://herschel.esac.esa.int/Docs/TechnicalNotes/
HIFI-Beam-Efficiencies-17Nov2010.pdf)

\begin{table}
\caption{O-rich AGB stars in the sample. Indicated is the
number of frequency settings observed for each object.}% The quoted 
%distance and mass-loss
%rates are taken from (a) \cite{debeck10}; (b) \cite{maercker09}
%and (c) \cite{just06}}
\label{sources}
\begin{tabular}{llllcc}
\hline \hline
  name     &    RA        &     Dec & obs & D & $\dot{M}$ \\
           &              &      &  & (pc)  &($M_{\odot}$\,yr$^{-1}$)\\
\hline
IRC+10011  & 01 06 26.0  & +12 35 53.0 & 15 & 740 & 1.9E-5$^{(1)}$\\
{\it o} Cet & 02 19 20.8 & $-$02 58 39.5 & 9  & 107 & 2.5E-7$^{(1)}$\\
IK Tau     & 03 53 28.9  & +11 24 21.7 & 15 & 260 & 4.5E-6$^{(1)}$\\
R Dor      & 04 36 45.6  & $-$62 04 37.8 & 9  & 59  & 2.0E-7$^{(2)}$\\
TX Cam     & 05 00 50.4  & +56 10 52.6 & 9  & 380 & 6.5E-6$^{(1)}$\\
W Hya      & 13 49 02.0  & $-$28 22 03.5 & 16 & 77  & 7.8E-8$^{(1)}$\\
AFGL~5379  & 17 44 24.0  & $-$31 55 35.5 & 8  & 580 & 2.0E-4$^{(3)}$\\
OH~26.5+0.6 & 18 37 32.5 & $-$05 23 59.2 & 8  & 1370& 2.6E-4$^{(3)}$\\
R Cas      & 23 58 24.9  & +51 23 19.7 & 9  & 106 & 4.0E-7$^{(1)}$\\
\hline
\end{tabular}
\tablebib{The quoted distance and mass-loss
rates are taken from (1) \cite{debeck10}; (2) \cite{maercker09};
and (3) \cite{just06}
}
\end{table}

In three objects, IRC+10011, IK Tau, and W Hya, we observe a large
number of frequency settings (15, 15 and 16, respectively),
with the intention of using them as
templates for other objects with high, intermediate and low mass-loss 
rates, respectively. In order to study a larger sample of stars,
we also observed 6 further
objects with fewer frequency settings selected from the original
settings (Table~\ref{sources})
which include many diagnostic lines (see Table~\ref{lines_id}).

In order to estimate the integrated line intensity, a Gaussian fit was 
performed in IDL.
%, which also give us the full-width at half maximum (FWHM) for each line. 
Some lines are flat-topped and hence the line intensity was
estimated using a rectangular (rhombic) profile. Due to the observing 
mode used, it is not possible to recover the true continuum level
of the spectra. 
However, the estimated line intensities are not affected by this
since we subtract the continuum prior to calculating them.
Our error estimates reflect only the noise in the 
baseline. Other uncertainties such as the assumption that a line
is Gaussian are not included. However, independent measurements of the
line intensity by summing the area under the line agree within twice the
estimated uncertainties listed in Table~\ref{lines_id}.
The absolute flux calibration error is expected to be $\sim$ 30\% in the
HIFI bands 6 and 7 and less ($\sim$ 15\%) at lower frequencies.

\section{Molecular lines}

We identify a total of 9 different molecular species 
as well as %their
associated isotopologues (Table~\ref{lines_id}). Most of the lines are
in the ground vibrational state. However, for H$_{2}$O and SiO, vibrationally
excited lines are also seen indicating that the lines originate from 
the hotter part of the envelope (Sect.~\ref{sec_vexp}). 
Many of the settings observed in the
star IK Tau have been presented
by \cite{decin10h} and are included here for completeness of our
sample, along with new observations.

It has been postulated that H$_{2}$O is one of the main molecular coolants
in the circumstellar envelopes of O-rich AGB stars, along with CO 
\citep{gold76}. This was confirmed by observations by
the Infrared Space Observatory (ISO) of
these stars which show numerous strong lines of both these molecules
(e.g. \citealt{barlow96, neufeld96}). We detected a total of
23 H$_{2}$O lines in all the three main isotopologues covering the
excitation temperature from 30 to 2350\,K. These lines probe the
full extent of the circumstellar envelope and due to HIFI resolution,
the lines are well resolved, enabling us to study the dynamics of
the dust-driven wind. It should be noted that the 
H$_{2}$O $1_{1,0}$-1$_{0,1}$ vibrationally excited line at 658.0\,GHz is 
likely to be a maser, as well as the 620.7\,GHz 
(5$_{3,2}$-4$_{4,1}$) line
(see Sect.~\ref{sec_maser}). This latter line has previously been reported
in the supergiant VY CMa \citep{harwit10}.
As for CO, we detected a total of 8 transitions in three isotopologues.

Other  molecules are also identified, such as NH$_{3}$,
SiO, HCN, SO, SO$_{2}$ and OH. These molecules have already been
detected in HIFI spectra of AGB and post-AGB stars \citep{bujar10, 
decin10h, menten10, just10, schoier11}.
%For SiO, we detected the vibrationally excited lines
%in a number of our objects which originate close to the dust condensation
%radius (Sect.~\ref{sec_vexp}). 
In AFGL~5379, a line is detected at 1196.010\,GHz which can be
attributed to the H$_{2}$S (3,1,2)-(2,2,1) transition. This line is not seen
in the other stars in our sample, however it is present in
the supergiant VY CMa (Alcolea 2011, in preparation).
A line is also seen in the spectrum of
OH~26.5+0.6 at 1114.431\,GHz which corresponds to the
vibrationally excited SiS J=62-61 transition. Although, with the high
upper energy level of 1918.3836\,cm$^{-1}$, we classify this line as
unlikely.

Since HIFI employs a double-side band mode, there is a possibility of 
ambiguity in classifying a line. The line at 1112.833\,GHz in the lower
side-band due to $^{29}$SiO J=26-25 coincides with the H$_{2}$O 
11$_{6,6}-12_{3,9}$ at 1101.130\,GHz. Considering the upper energy levels
and the expected line strengths of the two, we conclude that the line
is likely due to $^{29}$SiO.

\subsection{Line intensities}

\begin{figure}
\resizebox{\hsize}{!}{\includegraphics[width=7cm]{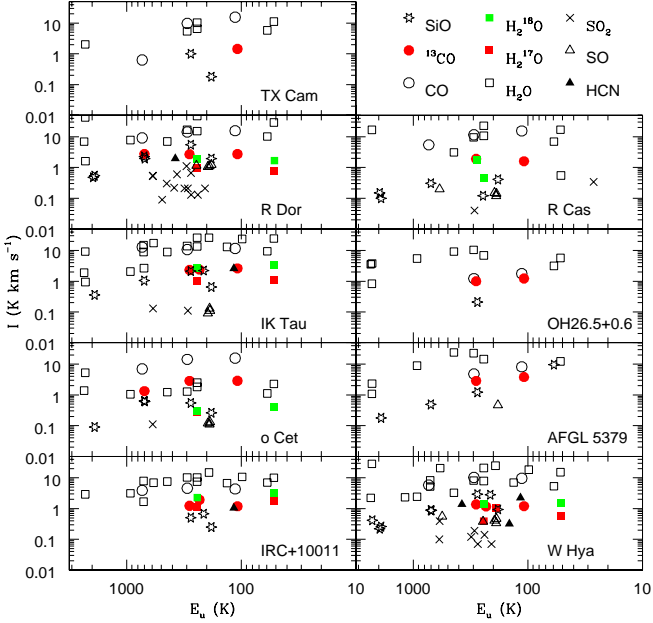}}
\caption{Integrated line intensities for all stars in the sample.
$^{12}$CO : black open circles; $^{13}$CO red filled circles; H$_{2}^{16}$O :
open squares; H$_{2}^{17}$O : red filled squares; H$_{2}^{18}$O : green 
filled squares; SiO : stars; SO : open
triangles; SO$_{2}$ : x; HCN : filled triangles.
}
\label{fig_flux}
\end{figure}

All objects show strong emission due to H$_{2}$O and CO molecules
(Fig.~\ref{fig_flux}).
The OH/IR stars,
OH~26.5+0.6, AFGL~5379 and IRC+10011, show very strong emission of both
ground state ortho- and para-H$_{2}$O lines compared to that of CO J=6-5.
Despite the high mass-loss rates in these OH/IR stars, the CO J=16-15
is very weak or not detected. This may be partly 
due to the attenuation of stellar light by dust, preventing the excitation
of this line.
%contributed to  the very strong dust continuum in these objects. 
In {\it o} Cet, the
CO lines are much stronger than other molecular species. This is
likely due to the white-dwarf companion with its hard X-ray flux
\citep{odwyer03}
photodissociating  H$_{2}$O
molecules in the circumstellar envelope of the primary star.
In all the stars, we
detect both $^{12}$CO and $^{13}$CO. The line ratios of the
corresponding transitions vary from 1.5 in stars with high mass-loss rate
to up to 10 in stars with lower mass-loss rates.

We detected both the ortho- and para-H$_{2}$O as well as the isotopologues
and vibrationally excited lines. A discussion 
on isotopic ratios of detected lines is presented by
Silva et al. (2011, in preparation).
A detailed modelling of radiative transfer 
of stars with low mass-loss rates
is in progress (Maercker et al. 2011, in preparation). The highest excitation
lines seen in our spectra are the vibrationally excited
$2_{12}-1_{01}$ and the vibrationally excited ground state lines of both
ortho- and para-H$_{2}$O with excitation energies in excess of
2000\,K (Table~\ref{lines_id}). These lines are likely radiatively pumped.

\begin{figure}
\resizebox{\hsize}{!}{\includegraphics[width=5cm]{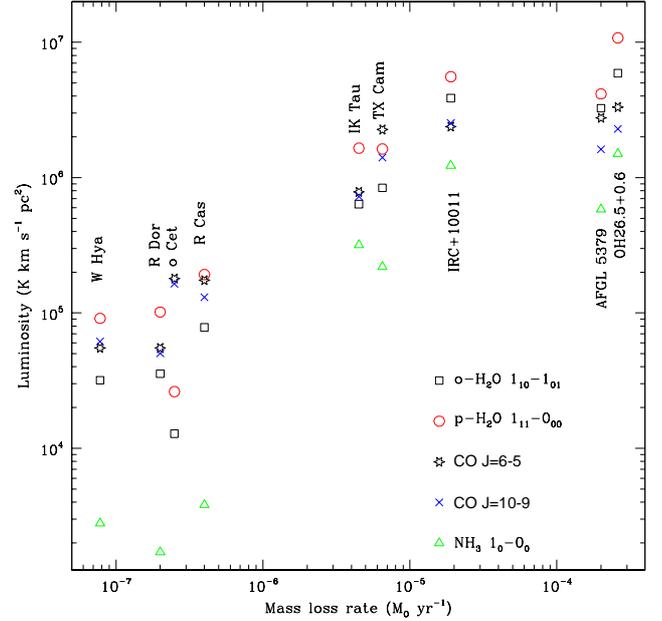}}
\caption{A plot of the line luminosity 
(squares : ortho-H$_{2}$O 1$_{1,0}$-1$_{0,1}$, circles : 
para-H$_{2}$O 1$_{1,1}$-0$_{0,0}$, stars : CO J=6-5,
x : CO J=10-9, triangles : NH$_{3}$ 1$_{0}$-0$_{0}$) as a function of the
published mass-loss rate show a strong correlation between
the two parameters.
}
\label{fig_linelum}
\end{figure}

%There is a strong positive correlation between the
%H$_{2}$O and CO line luminosity and the mass-loss rate.
In Fig.~\ref{fig_linelum}, we plot the line luminosity as a
function of mass-loss rate for CO J=6-5, 10-9, and
the ground-state lines of both ortho- and para-H$_{2}$O,
as well as for NH$_{3}$ 1$_{0}$-0$_{0}$.
The distances and mass-loss rates are taken from
\cite{debeck10, maercker09, just06}
(see Table~\ref{sources}). 
%In Fig.~\ref{fig_linelum}, we plot CO J=6-5, 10-9, and
%the ground-state lines of both ortho- and para-H$_{2}$O.
%p-H$_{2}$O 1$_{1,1}$-0$_{0,0}$,%
%4$_{2,2}$-4$_{1,3}$, 
%o-H$_{2}$O 1$_{1,0}$-1$_{0,1}$,
% 3$_{2,1}$-3$_{1,2}$, 3$_{1,2}$-2$_{2,1}$, 3$_{1,3}$-3$_{0,3}$
%since these lines are detected in all our objects. 
%%There is a strong positive correlation between individual
%%H$_{2}$O and CO  line luminosities and mass-loss rate, except for {\it o} Cet
%%where H$_{2}$O is likely photodissociated by the binary.
There is a strong positive correlation between individual
H$_{2}$O and CO  line luminosities, 
i.e., the slopes for the
mass-loss rate up to $\sim 10^{-5}$~$M_{\odot}$ yr$^{-1}$ are betwen
0.7-0.8.
There is one exception regarding the H$_{2}$O luminosity in {\it o} Cet 
which is much lower than for the CO. This is likely due to H$_{2}$O being
photodissociated by the binary companion.
The flattening off of this relation at $\dot{M} \sim 3 \times 10^{-5}$
$M_{\odot}$ yr$^{-1}$ is consistent with relations by \cite{debeck10}.

Another molecule readily detected in circumstellar envelopes
of AGB stars is SiO \citep[e.g.,][]{bujar94, olofsson98}. Here, we identified
three silicon isotopes. The vibrationally excited lines
of high-J transitions are also seen. 

SO has been observed in a number of AGB stars covering a wide range 
in mass-loss rates. The estimated abundance is a few 10$^{-7}$ 
up to $10^{-6}$ $M_{\odot}$ yr$^{-1}$
\citep{omont93, bujar94}.
However, we only observed SO and SO$_{2}$ lines in our HIFI frequency range
in stars with low mass-loss rates. 
From ISO observations, SO$_{2}$ absorption is detected
towards stars with optically thin envelopes \citep{yamamura99},
while this is not seen in optically thick shells.
SO is thought to be formed mainly via S + CO and to a 
lesser extent via S + OH \citep{cherchneff06} and is
expected to be present throughout the envelope. It follows then that
the formation of SO$_{2}$ is via SO + OH. The best example for 
SO and SO$_{2}$ can be seen in our spectrum of R Dor 
(Fig.~\ref{fig_rdor}). In this object,
the line fluxes of the SO lines are about an
order of magnitude larger than that of  SO$_{2}$. In the other
stars where both molecules are observed, the flux ratios
are less than 5. The only other S-bearing molecule seen in our spectra
is H$_{2}$S in AFGL~5379 at 1196.0\,GHz. This line is absent in all 
the other stars in our sample, compare to a more ubiquitous
presence of this molecule observed from the ground in OH/IR stars
\citep{omont93}.

We detected the OH triplet line at 1834.7\,GHz, which is very strong
in the two extreme OH/IR stars, in agreement with the fact 
that these stars emit strongly in the OH 1612\,MHz maser 
\citep[e.g.,][]{sevenster01}. The spectrum of IRC+10011 is contaminated 
by ripples (standing waves in the HEB mixers) 
preventing us from confirming the presence of the OH line.

\begin{figure}
\resizebox{\hsize}{!}{\includegraphics[width=5cm]{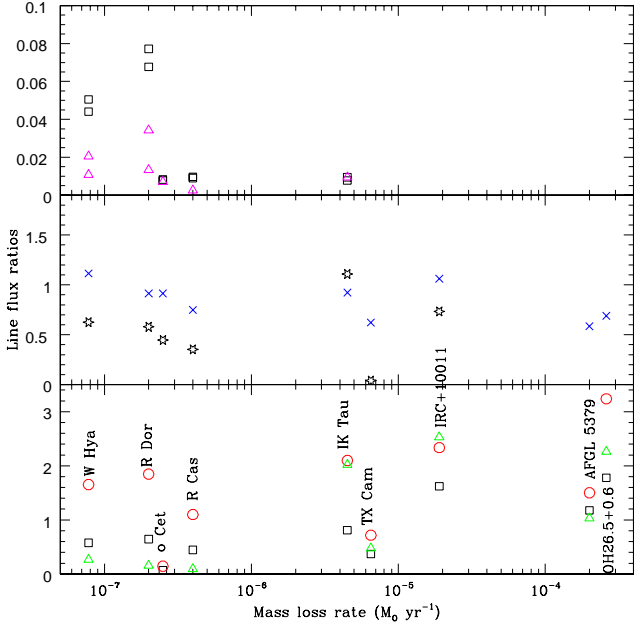}}
\caption{A plot of the line ratios relative to the integrated intensity of the
CO J=6-5 as a function of mass-loss rates. 
In the lower panel : for ortho-H$_{2}$O 1$_{1,0}$-1$_{0,1}$ (squares), 
para-H$_{2}$O 1$_{1,1}$-0$_{0,0}$ (circles), NH$_{3}$ 1$_{0}$-0$_{0}$
(triangles) scaled up by a factor of 5. The 
the middle panel shows CO J=10-9 (x), CO J=16-15 (stars) 
and in the upper panel : SO (13$_{13}$-12$_{12}$) 559.3 and 
SO (13$_{14}$-12$_{13}$) 560.2\,GHz (triangles) 
and SO$_{2}$ (21$_{6,16}$-21$_{5,17}$)
558.4 and SO$_{2}$ (37$_{1,37}$-36$_{0,36}$) 659.4\,GHz (squares).
}
\label{fig_ratio}
\end{figure}

To study the effects of mass-loss rate on the excitation, we plot
line intensity ratios of CO J=16-15 and J=10-9 relative to the J=6-5
(Fig.~\ref{fig_ratio} middle panel). These ratios appear to be 
constant over four orders of magnitude of mass-loss rates,
indicating that the density is higher than critical density of these 
transitions, i.e., the lines are thermalized in the 
emitting region. 
%the excitation of high-J CO lines is independent of the density.
The line ratios of the ground state lines
of both ortho- and para-H$_{2}$O with CO, on the other hand, 
show an increasing trend as
a function of mass-loss rate with the para-H$_{2}$O having a 
consistently higher ratio than that of the ortho-H$_{2}$O. 
This same increasing trend can be seen between the NH$_{3}$ 1$_{0}$-0$_{0}$
and mass-loss rates,
indicative of similar excitation conditions for both  molecules.
The top panel of Fig.~\ref{fig_ratio} shows the SO/CO and SO$_{2}$/CO ratios.
Here, a decreasing trend for the line ratio with the mass-loss rate can
be seen. From the HIFI observations, both SO and SO$_{2}$ are not detected 
in stars with a mass-loss rate higher than 10$^{-5}$ $M_{\odot}$ yr$^{-1}$. 
It is therefore not possible to
confirm the tight correlation seen by \cite{olofsson98} between 
CO (J=1-0) and SO (J$_{\rm k}$=3$_{2}$-2$_{1}$) line fluxes. 
\cite{omont93}, however, detected
low excitation lines of SO and SO$_{2}$ (T$_{\rm ex} \leq$ 55\,K) in 14
OH/IR stars, including three stars in our sample (IRC+10011, IK Tau and
OH~26.5+0.6). In optically thick envelopes, the high excitation levels
of the sulphur-bearing molecules in the HIFI range (T$_{\rm ex} \geq$ 200\,K) 
are simply not sufficiently excited.

\subsection{H$_{2}$O masers}
\label{sec_maser}

\begin{figure}
\resizebox{\hsize}{!}{\includegraphics[width=7cm]{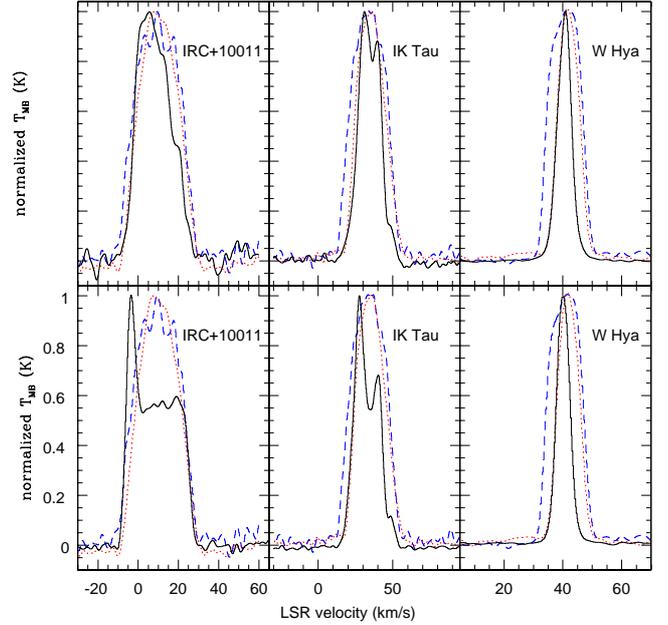}}
\caption{The lower panels show
a comparison of the normalized line profile of the
620.7\,GHz H$_{2}$O 5$_{3,2}$-4$_{4,1}$
line (solid line) with the ground state ortho-H$_{2}$O 1$_{1,0}$-1$_{0,1}$
line at 556.9\,GHz (dotted red line) and the CO J=6-5 (dashed blue line).
The upper panels show the 970.3\,GHz H$_{2}$O 5$_{2,4}$-4$_{3,1}$ (solid) 
with same the H$_{2}$O ground state and CO lines.
}
\label{fig_maser620}
\end{figure}

Maser emission has been observed toward a large number of
evolved stars. Its signature includes an anomalously strong
line intensity and a narrow line width due to maser amplification. 
In some instances, a maser can be seen in a narrow peak in the blue wing.
%One of the strongest
%masers observed is the 1612~MHz OH maser \citep[e.g.,][]{sevenster01}.

In three objects (W Hya, IK Tau, and IRC+10011), we observe the
620.7\,GHz transition of ortho-H$_{2}$O 5$_{3,2}$-4$_{4,1}$ which is
predicted to be masing \citep{neufeld91}. The maser line is
brighter compared to the thermal excited lines and tends to have
a narrower line profile as the region of coherent velocity
required to produce maser emission is generally small.
In W Hya, the maser line is narrow compared to the thermally excited
lines of H$_{2}$O 1$_{1,0}-1_{0,1}$ and 
CO J=6-5, indicating it comes from a region close to the star.
In IK Tau, the line exhibits a double-peak profile which has a line width
slightly narrower than that of the 556.9\,GHz ground-state ortho-H$_{2}$O 
and the CO line.
In IRC+10011, however, the line is of the same width as the ground
state H$_{2}$O and
CO J=6-5 lines, and is thought to arise from thermal emission,
but it also has a strong blue component due to the masing action
(see Fig.~\ref{fig_maser620} lower panels).
This is in line with the expectation that the 620.7\,GHz line is masing at high 
($\sim 10^{3}$\,K) temperature \citep{neufeld91}.
The maser emission appears to arise from the blue-shifted part of the 
spectrum, i.e., in the front side of the envelope.
 \cite{neufeld91} also predicted a maser line at 970.3\,GHz from
para-H$_{2}$O (5$_{2,4}$-4$_{3,1}$) which we observed in W Hya and IK Tau
with similar line widths and profiles as for the 620.7\,GHz maser.
In IRC+10011, we did not detect a maser peak (Fig.~\ref{fig_maser620} 
upper panels), and the line is about the same brightness as for 
the ground transition line at 556.9\,GHz (Table~\ref{lines_id}), 
hence the maser action appears to be quenched in this object.

\begin{figure}
\resizebox{\hsize}{!}{\includegraphics[width=7cm]{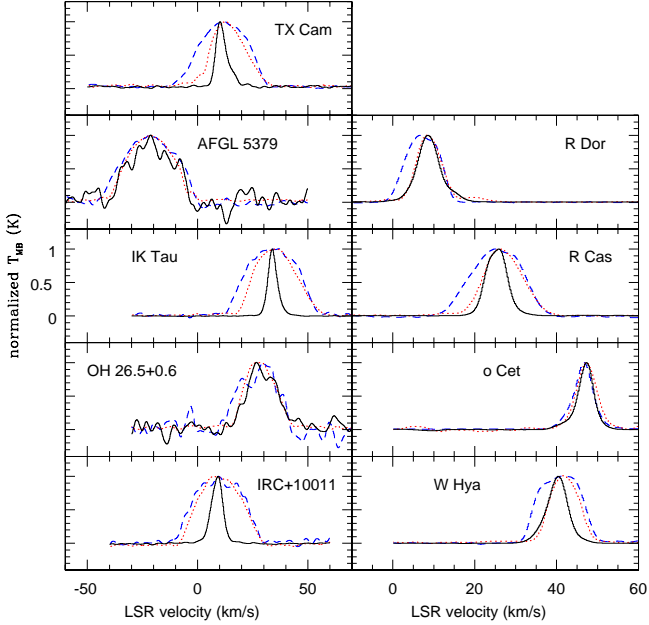}}
\caption{A comparison of the normalized line profile of the
658.0\,GHz H$_{2}$O line (solid line) with the ground state ortho-H$_{2}$O
line at 556.9\,GHz (dotted red line) and the CO J=6-5 (dashed blue line).
}
\label{fig_maser658}
\end{figure}

In one of the frequency settings, we detected the vibrationally excited
H$_{2}$O $1_{1,0}$-1$_{0,1}$ line at 658.0\,GHz which is likely masing
(Fig.~\ref{fig_maser658}). In all the objects with low and intermediate
mass-loss rates, the maser line is markedly narrow compared to the
556.9\,GHz line %and occur close to the blue side of the spectrum, 
except for {\it o} Cet, where both lines have comparable widths.
This indicates, again, that the maser line in the majority of the stars
originates close to the star where the wind has not yet reached the 
terminal velocity. The intrinsic line intensity %of these stars
supports the fact that the line is masing, i.e., the line intensity
of the vibrationally excited line is very bright, compared with
the ground state line at 556.9\,GHz. However, in extreme OH/IR stars
(OH~26.5+0.6 and AFGL~5379),
the widths of the ground and vibrationally excited lines are
comparable and the flux of the vibrationally excited line is 
small relative to the ground state line. It is unlikely then
in these two objects that the line is masing. This is probably due to
the high density quenching maser action in the inner part of 
the envelope.

We also observe the vibrationally excited line of 
para-H$_{2}$O 1$_{1,1}$-0$_{0,0}$
at 1205.8\,GHz. Unlike the vibrationally excited 1$_{1,0}$-1$_{0,1}$
ortho-H$_{2}$O, this line does 
not appear to be masing when comparing with the same transition in 
the ground vibrational state at 
1113.3\,GHz as the vibrationally excited line is much weaker. 

\subsection{Expansion velocity of circumstellar envelops}
\label{sec_vexp}

\begin{figure*}
\sidecaption
\includegraphics[width=12cm]{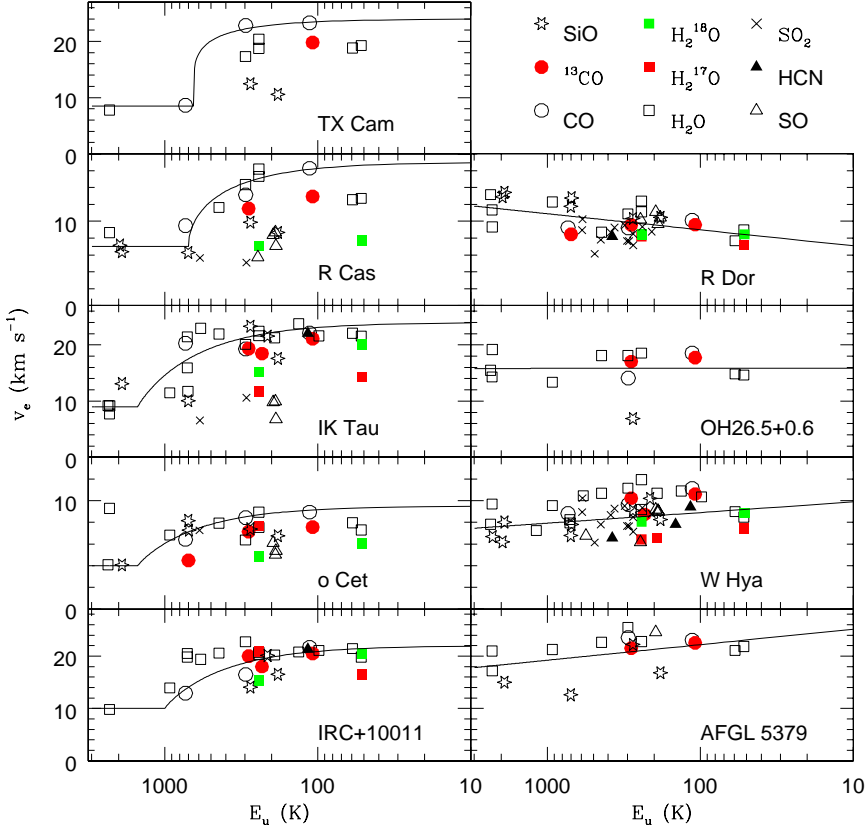}
\caption{Observed expansion velocity for molecular lines in our sample stars.
}
\label{fig_vexp}
\end{figure*}

Once dust grains condense in the outflow of AGB stars,
the wind acceleration mechanism is thought to be due to radiation
pressure on the dust grains, which drags the gas as they
move away from the central star (e.g., \citealt{gold76,habing94}). 
Lines with a high excitation temperature are expected to be emitted
in regions close to the star, where the wind has not yet reached its 
terminal velocity. Consequently, these lines would be observed to be 
narrower than lines with lower excitation temperatures, emitted
from farther out in the envelope.
\cite{just94} demonstrated that in a radiative transfer calculation
for successively higher-J CO lines, 
the warmer regions of the emitting zone are being probed which have
smaller line widths than for the lowest CO transition. However, 
observations of maser lines which probe regions close to the central star
revealed that the wind acceleration is not as fast as predicted
by the dust-drag wind \citep{chapman94}. In Fig.~\ref{fig_vexp},
we plot the expansion velocities for 
the lines detected in our spectra, which probe the warm region as
well as the region close to the central star. 
%These are the vibrationally excited H$_{2}$O and SiO lines. 
Here, the observed expansion velocity is a measure of half the width 
at the baseline level. The uncertainty of the expansion velocity
is $\sim$ 15-30\%. In most cases,
there is a trend that lower excitation lines have wider observed line 
profiles, indicating they come from regions where the wind is close to the
previously determined expansion velocity from ground-based observations
of low-J CO lines. This indicates that there is a velocity gradient in
the circumstellar outflow. One caveat is that for H$_{2}$O lines, the
profile can be heavily absorbed in the blue wing, leading to slightly
lower measured expansion velocities. This can be seen in 
Fig.~\ref{fig_vexp} which shows that the
measured expansion velocity of CO J=6-5 is larger than that for the ground
state of both ortho- and para-H$_{2}$O.
We observe isotopic lines of H$_{2}$O, and a trend can be seen that 
$\rm{v}(H_{2}^{16}O) > \rm{v}(H_{2}^{17}O,H_{2}^{18}$O).
%This points to the surprising fact that the less abundant species
%of H$_{2}$O suffer more from the blue wing absorption.
This may be due to the fact that the less abundant species are more prone to
photodissociation from the interstellar radiation hence we see molecules
close to the central star where the gas has not yet reached the 
terminal velocity.

For all stars, 3 CO lines have been targeted : J=6-5, 10-9 and 16-15.
The widths of these lines show the trend of increasing 
velocity for the lower excitation line, in agreement with the
acceleration. However, the J=16-15 line is not seen towards
the two extreme OH/IR stars, OH~26.5+0.6 and AFGL~5379, possibly due to
the dust attenuation of stellar radiation, preventing the excitation
of this line. We readily detected the $^{13}$CO lines in our frequency
settings as well and in most cases, the line widths of these two
isotopes are comparable.

The line widths of SO and 
SO$_{2}$  %for the lower energy 
are narrower than those from CO and H$_{2}$O. 
Due to the low abundance, the lines are not thought to be
optically thick, hence the reason why the lines are narrower is 
likely because they originate from the inner region of the envelope.
It can be that these molecules 
are also destroyed closer
in the central star by interstellar UV radiation, compared to
the more robust CO molecules. 
In some stars with a relatively low mass-loss rate,
the measured velocities of SiO J=16-15 and $^{29}$SiO J=13-12
lines also are smaller than those from CO.
%It can be that these molecules are also destroyed closer
%in the central star by interstellar UV radiation, compared to
%the more robust CO molecules. 

A number of stars show a clear trend of the high excitation lines
having a smaller line width than low excitation ones 
(Fig.~\ref{fig_vexp}), consistent with a region 
where the molecules are then accelerated towards the
terminal velocity observed in the ground-based low-J CO observations.
In order to qualitatively show that the wind is accelerated,
we overplot a function similar to the dust-driven wind of \cite{lamers99}
but with a dependence on the energy rather than the radius :
\begin{equation}
\rm v = \rm v_{0} + (\rm v_{e} - \rm v_{0}) [1. - E_{0}/E]^\beta
\end{equation}
where $\rm v_{0}, \rm v_{e}$ and $E_{0}$ are arbitrary initial 
and expansion velocities in km s$^{-1}$
and energy in K, respectively, and $\beta$ is a velocity exponent
(Table~\ref{vfit}).
The parameters are chosen to follow the $^{12}$CO transitions as a 
the main tracer for the expansion velocity.
In a few cases, however, 
no clear jump in velocity between high
and low excitation lines is seen,
and a straight line can better describe 
the velocity field (right panels in Fig.~\ref{fig_vexp}). 
The fitted lines are least-square-fits to all the data points.
In general, despite the smaller observed expansion velocity of
SO, SO$_{2}$, SiO and less abundant isotopologues of H$_{2}$O and CO,
the measured velocities of these species follow the overall trend of
%the velocity field in our sample stars show the overall trend of 
increasing velocity with decreasing 
excitation energy, i.e., increasing radius from the central star.

\begin{table}
\caption{Parameters describing the velocity profile
for stars which show wind acceleration (Fig.~\ref{fig_vexp}).
}
\label{vfit}
\begin{tabular}{lcccc}
\hline
           & $\rm v_{0}$ & $\rm v_{e}$ & $E_{0}$ & $\beta$ \\
         &(km s$^{-1}$)&(km s$^{-1}$)& K &    \\
\hline \hline
TX Cam     & 8.5    & 24.0    & 650.   & 0.2 \\
R Cas      & 7.0    & 17.0    & 700.   & 0.5 \\
IK Tau     & 9.0    & 24.0    & 1500.  & 0.9 \\
{\it o} Cet & 4.0    & 9.5     & 1500.  & 0.9 \\
IRC+10011  & 10.0   & 22.0    & 1000.  & 0.9 \\
\hline
\end{tabular}
\end{table}
%}
In the case of R Dor, however, the highly excited lines of both H$_{2}$O  
and SiO indicate a larger expansion velocity in the inner part of the 
envelope than for the lower J-transitions of CO. 
This is counter-intuitive compared to what is expected of an 
accelerating dust-driven wind and may pose a challenge to detailed
radiative transfer modelling of the stellar wind of R Dor.
Also in AFGL~5379 and OH~26.5+0.6, the expansion 
velocity of various lines suggests that the wind reaches the terminal
velocity already in the innermost part, i.e., the acceleration zone
is very small since the line widths are similar regardless of the
excitation. For W Hya, the line widths increase linearly with the
decreasing excitation energy. These stars demonstrate that the
gas-dust interaction does not seem to follow a simple momentum
transfer previously used to describe the wind. To fully understand
the observed line widths, a full radiative transfer must be 
performed including a velocity field and temperature profile.

\section{Summary}

We present the full dataset on O-rich AGB stars as part of the
guaranteed time key program HIFISTARS to study the late stages of
stellar evolution both kinematically and dynamically.
We detect emission of 9 molecular 
species and their most common isotopologues. We find a trend of
increasing line luminosity of H$_{2}$O, CO and NH$_{3}$ with the mass-loss
rate (Fig.~\ref{fig_linelum}) which tapers off for stars with a 
high mass-loss rate. Interestingly, the line ratios of high-J CO relative 
to CO J=6-5 are independent of the mass-loss rate (Fig.~\ref{fig_ratio}).
This implies that the excitation of high-J CO is independent of the density,
i.e., the lines are thermalized.
Other species such as H$_{2}$O and NH$_{3}$ show a positive correlation
with the mass-loss rate.

From the line brightness and shape, we conclude that the H$_{2}$O
620.7\,GHz is masing in all three objects observed. The 970.3\,GHz and the
658.0\,GHz lines are masing in objects with relatively low 
($\leq 10^{-5}$ $M_{\odot}$ yr$^{-1}$) mass-loss rates. These maser lines
are generally significantly narrower than the thermal H$_{2}$O lines.

Since the lines are well resolved, we can use the observed line width to 
derive the expansion velocity for each transition. In general, highly
excited lines which originate close to the star have smaller expansion
velocities compared to the lower excitation lines (Fig.~\ref{fig_vexp}). 
This is in agreement with 
the gas being dragged by dust grains as they are accelerated outward
due to stellar radiation. However, from our observations, the acceleration
zone differs in our sample, with stars with a high mass-loss rate
showing high-excitation lines having velocity close to the terminal
velocity while stars with a low mass-loss rate having a distinct low
velocity region (E$_{u} \leq 10^{3}$\,K). In R Dor, the highly excited
lines show smaller observed line widths than the low excited lines. For this 
object, the gas appears to be decelerating as it moves away from the
central star. A caveat is the optical depth which can affect the line
profile. In order to calculate this effect, a full radiative transfer
has to be performed in individual objects. This work has started and will
be presented in the forth coming papers.

\appendix

\section{Spectra}

In this section, we present the observed spectra of O-rich
AGB stars observed in the HIFISTARS guaranteed time program,
along with the measured line intensities.

\begin{figure*}
\resizebox{\hsize}{!}{\includegraphics[width=7cm]{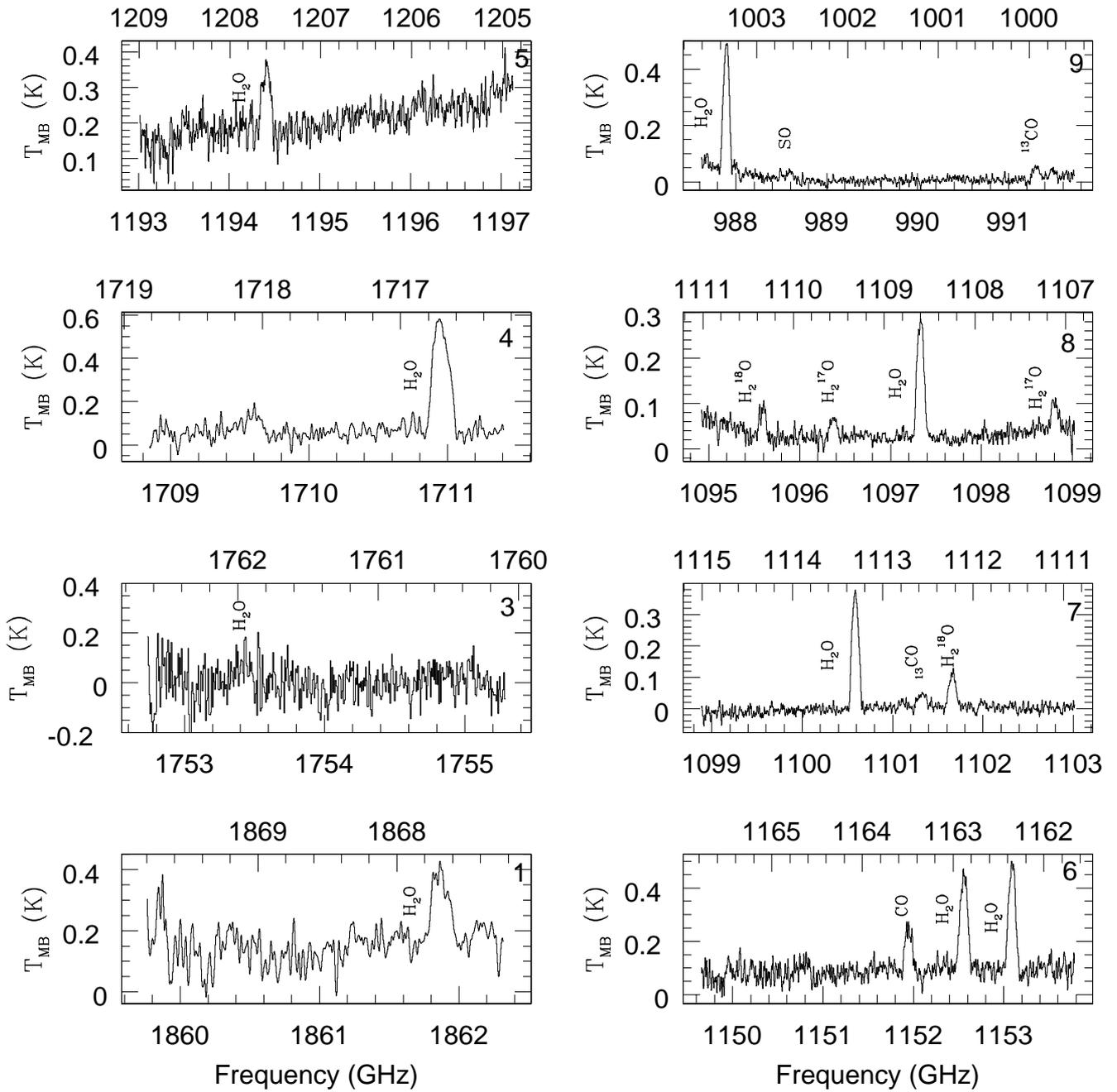}}
\caption{HIFI spectra of IRC+10011. For details of the lines detected, see
Table~\ref{lines_id}.
}
\label{fig_spec}
\end{figure*}

\begin{figure*}
\resizebox{\hsize}{!}{\includegraphics[width=7cm]{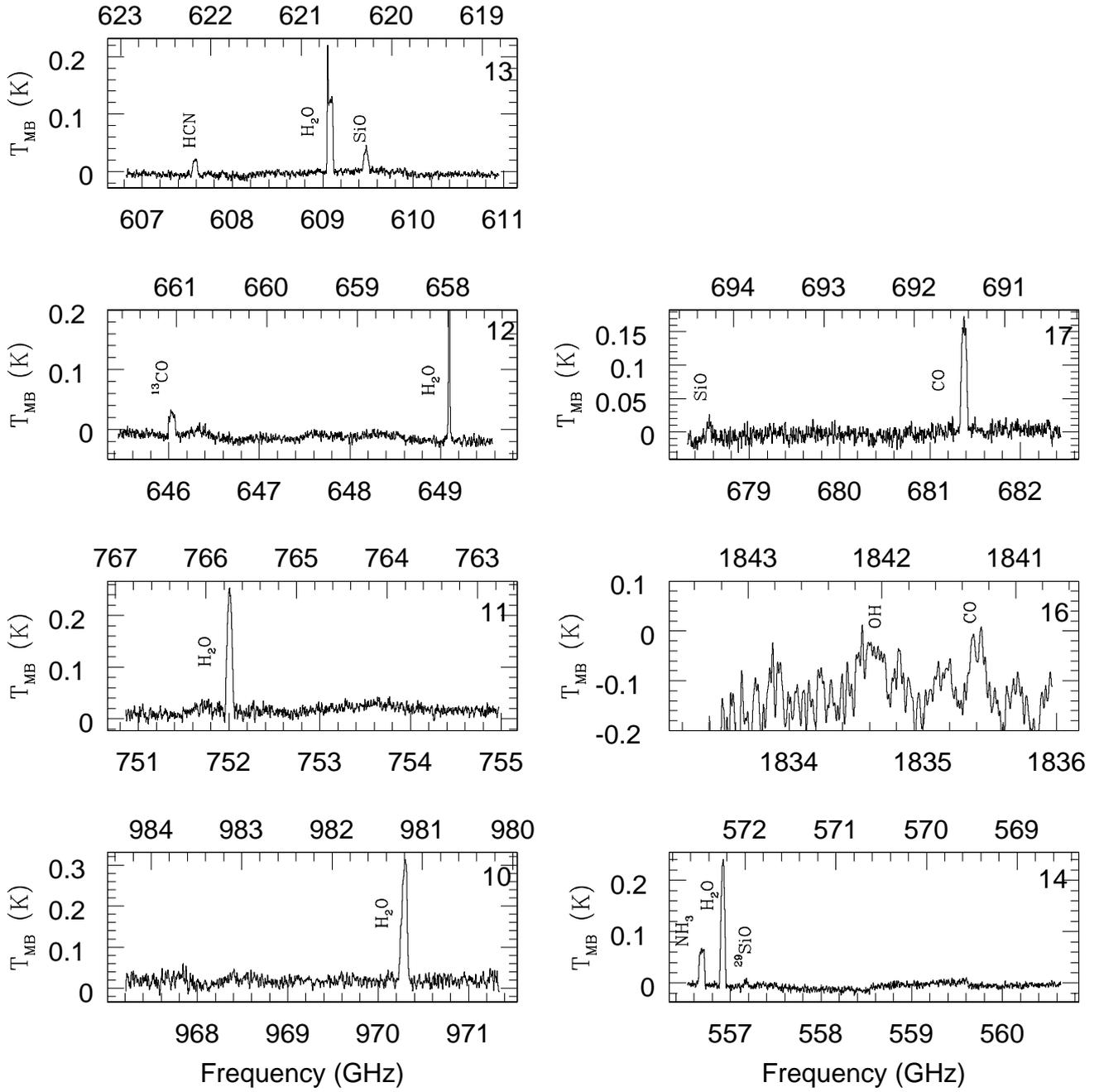}}
\caption{HIFI spectra of IRC+10011, cont.
}
\end{figure*}

\begin{figure*}
\resizebox{\hsize}{!}{\includegraphics[width=7cm]{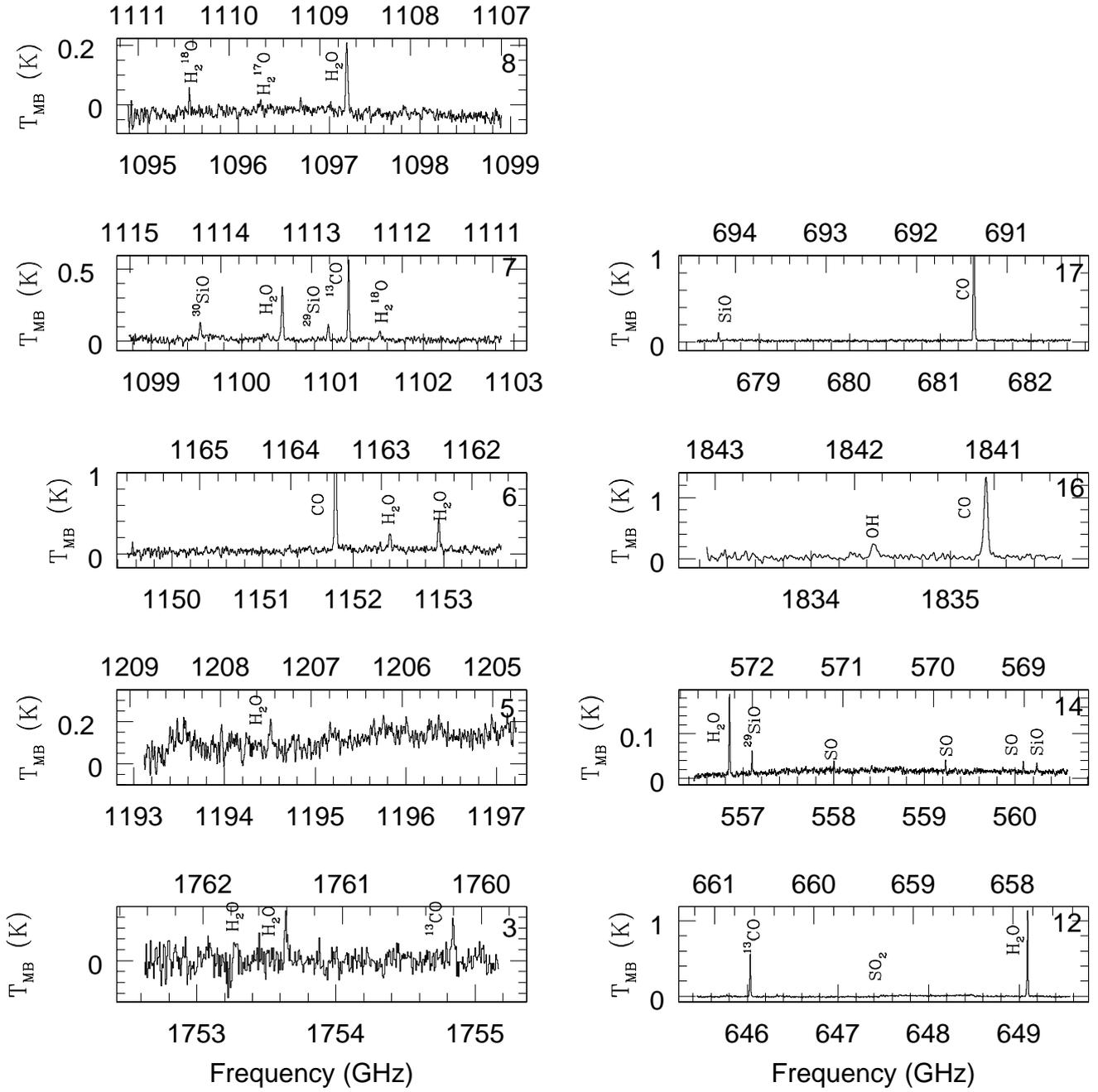}}
\caption{HIFI spectra of \it{o} Cet.
}
\end{figure*}

\begin{figure*}
\resizebox{\hsize}{!}{\includegraphics[width=7cm]{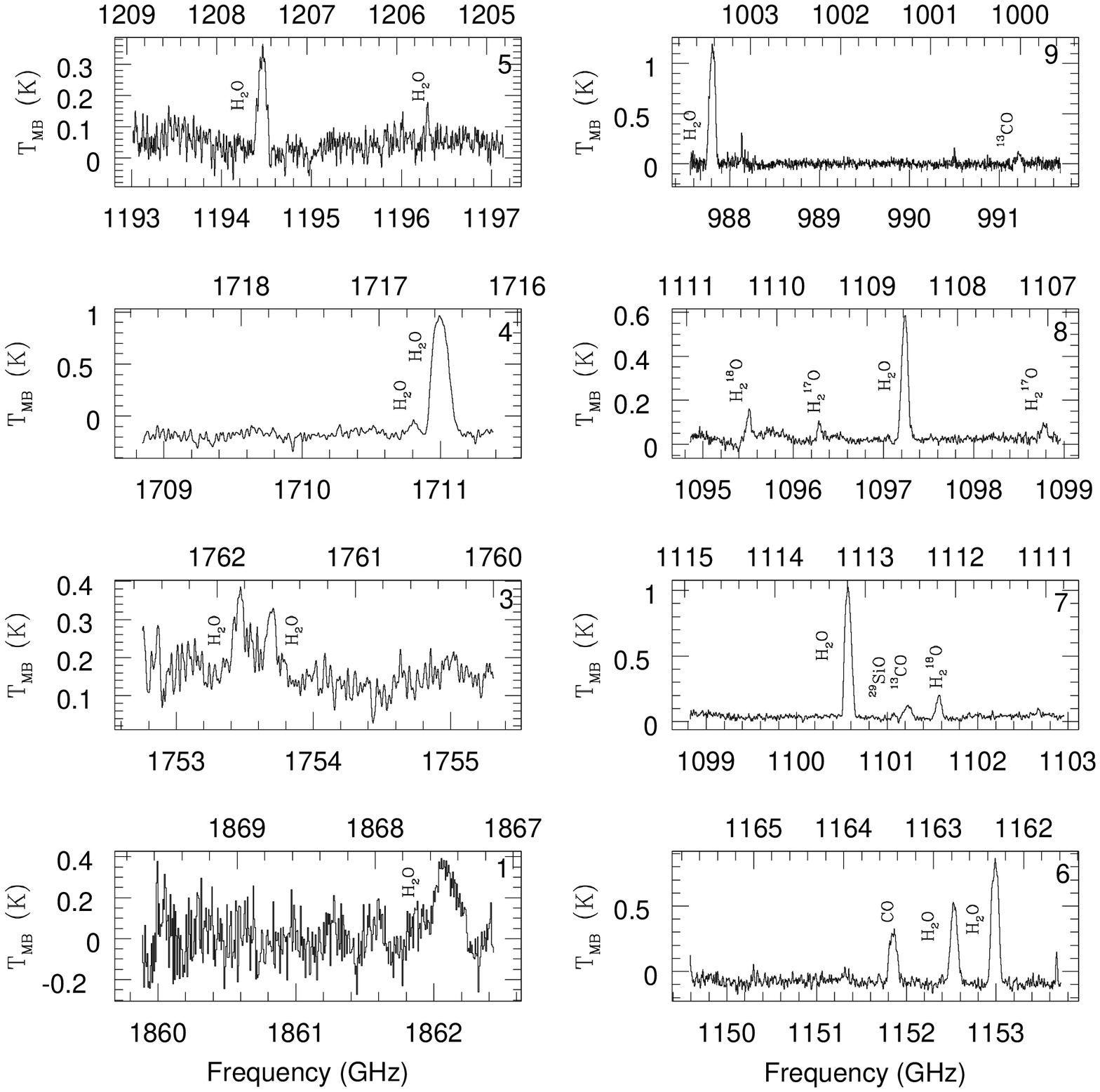}}
\caption{HIFI spectra of IK Tau.
}
\end{figure*}

\begin{figure*}
\resizebox{\hsize}{!}{\includegraphics[width=7cm]{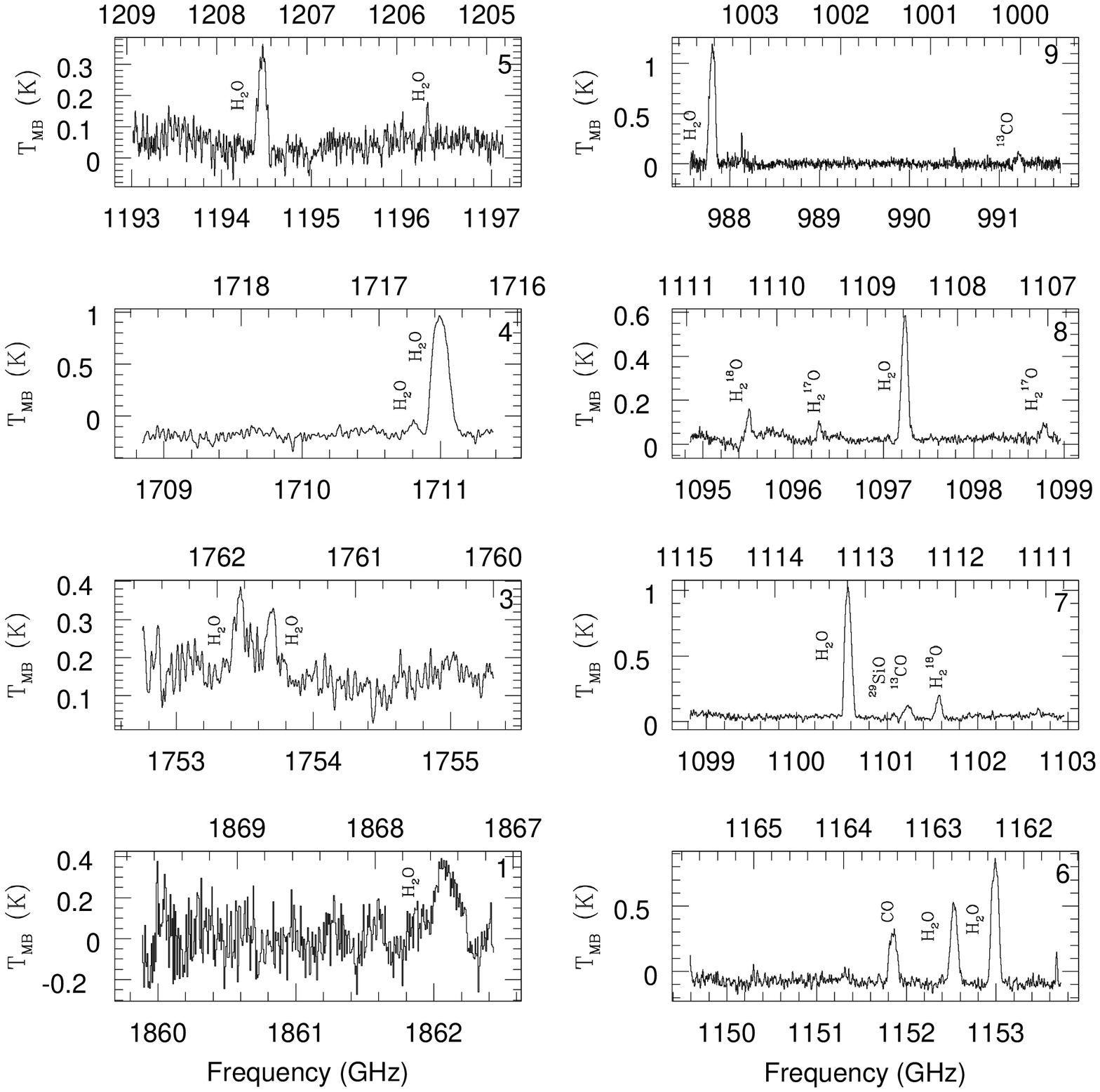}}
\caption{HIFI spectra of IK Tau, cont.
}
\end{figure*}

\begin{figure*}
\resizebox{\hsize}{!}{\includegraphics[width=7cm]{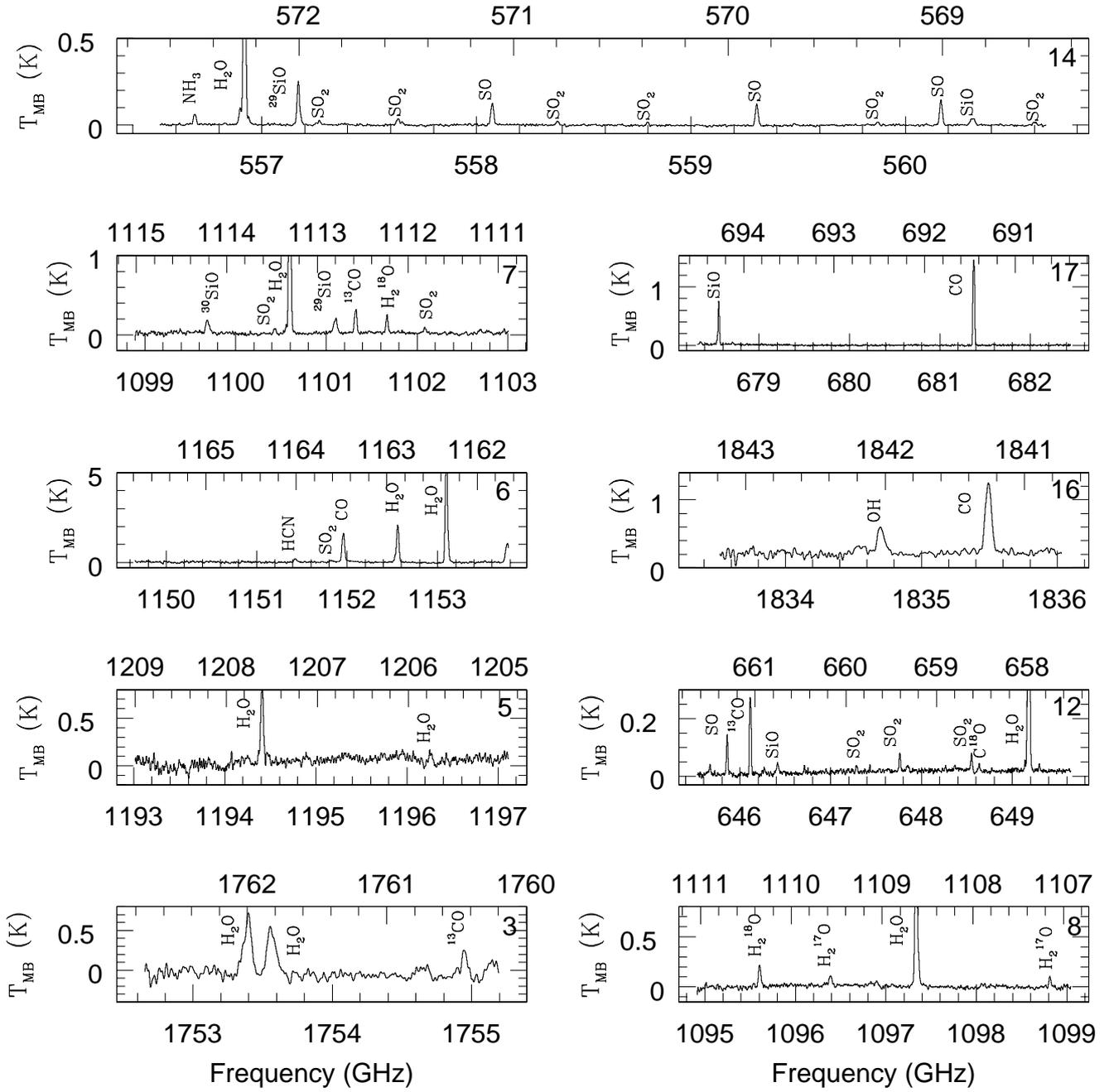}}
\caption{HIFI spectra of R Dor.
}
\label{fig_rdor}
\end{figure*}

\begin{figure*}
\resizebox{\hsize}{!}{\includegraphics[width=7cm]{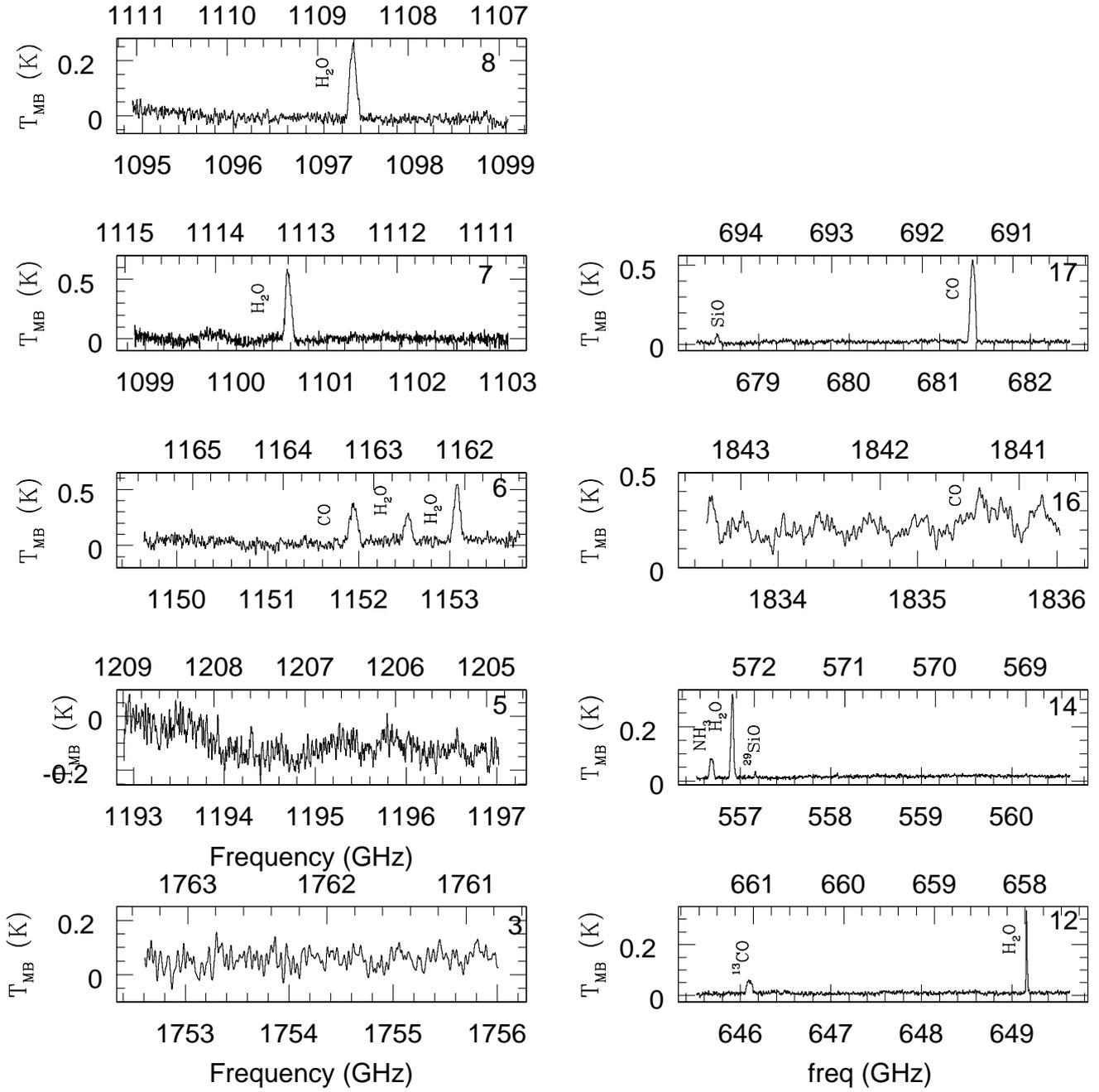}}
\caption{HIFI spectra of TX Cam.
}
\end{figure*}

\begin{figure*}
\resizebox{\hsize}{!}{\includegraphics[width=7cm]{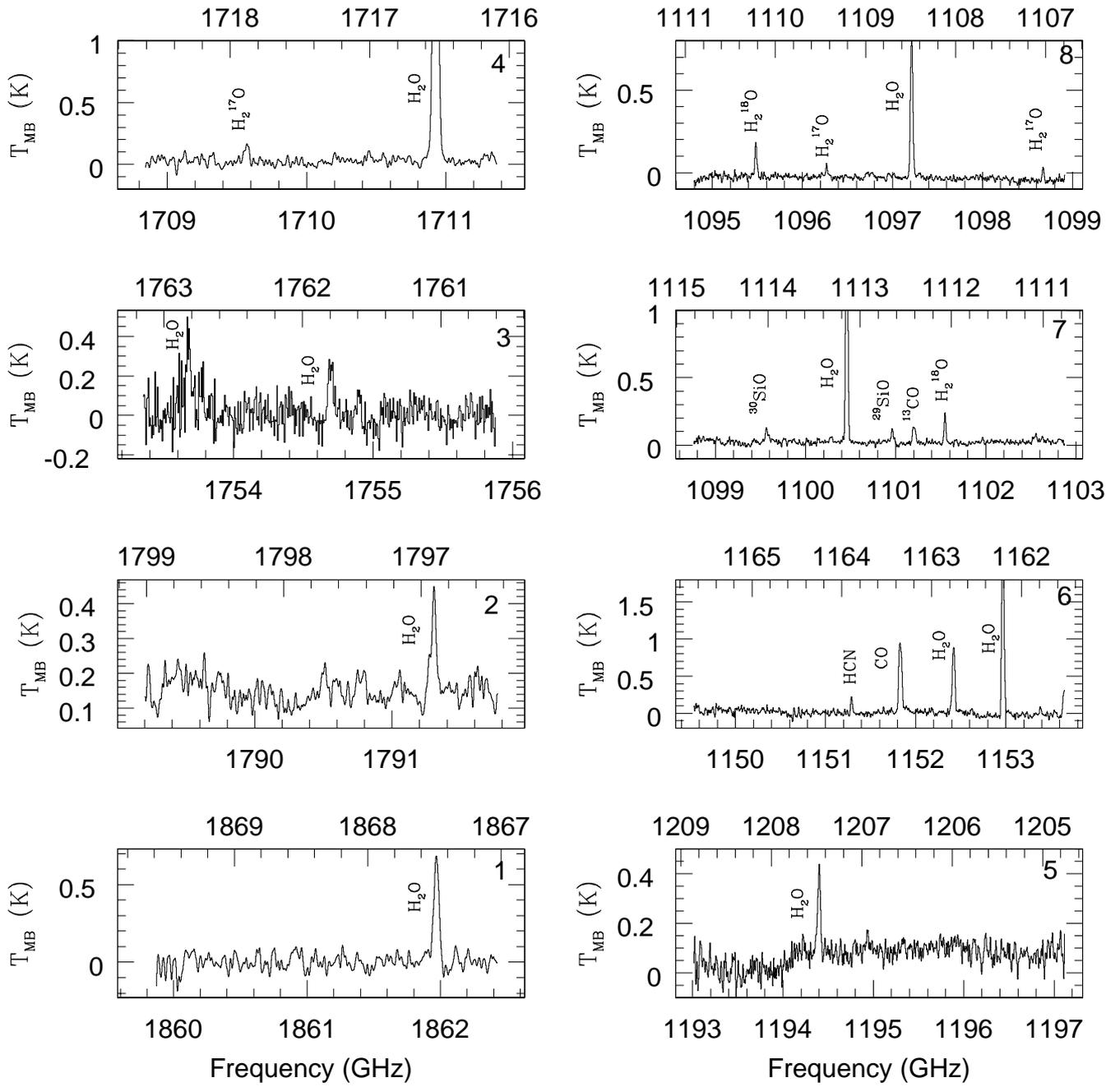}}
\caption{HIFI spectra of WHya.
}
\end{figure*}

\begin{figure*}
\resizebox{\hsize}{!}{\includegraphics[width=7cm]{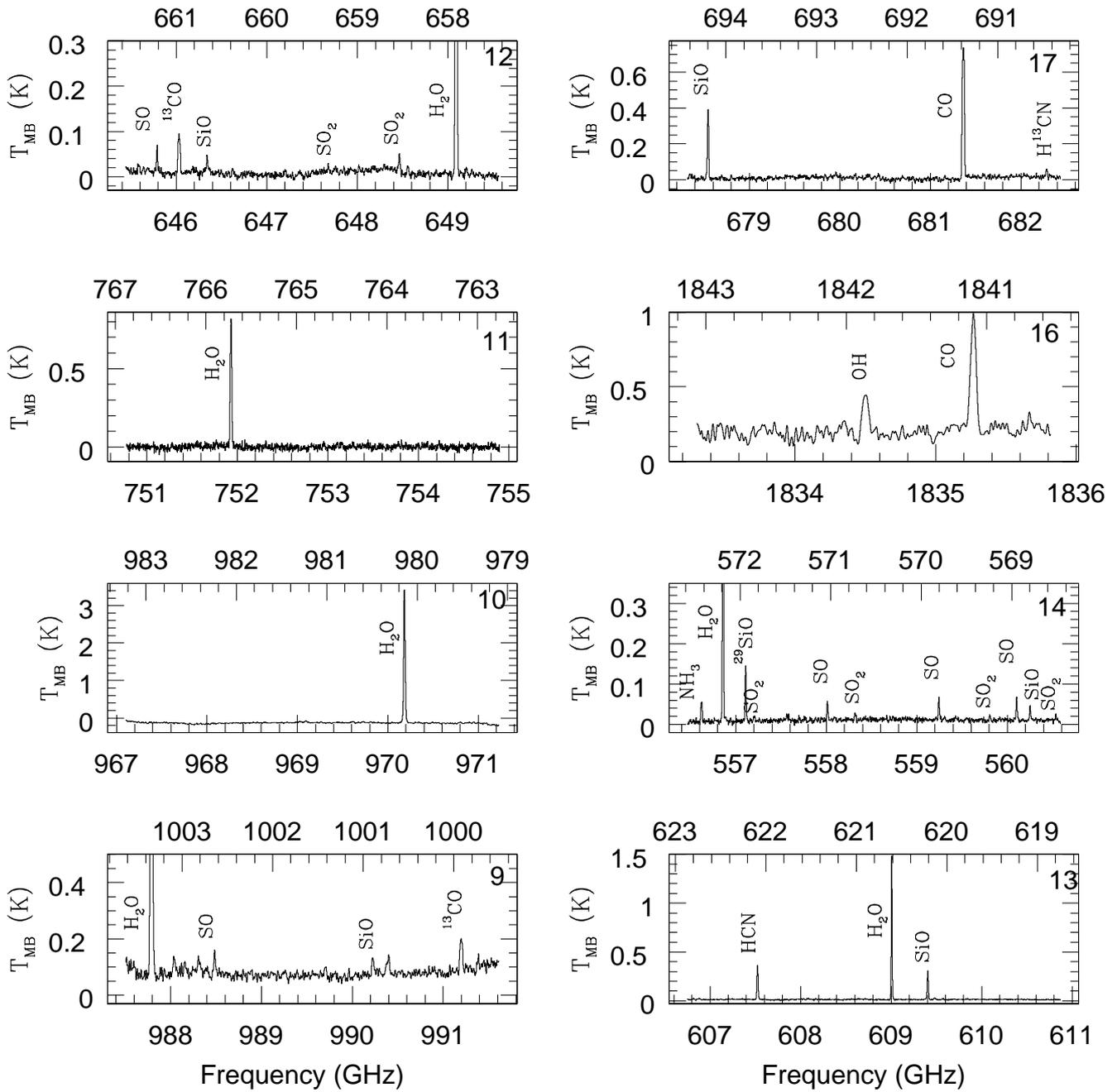}}
\caption{HIFI spectra of WHya, cont.
}
\end{figure*}

\begin{figure*}
\resizebox{\hsize}{!}{\includegraphics[width=7cm]{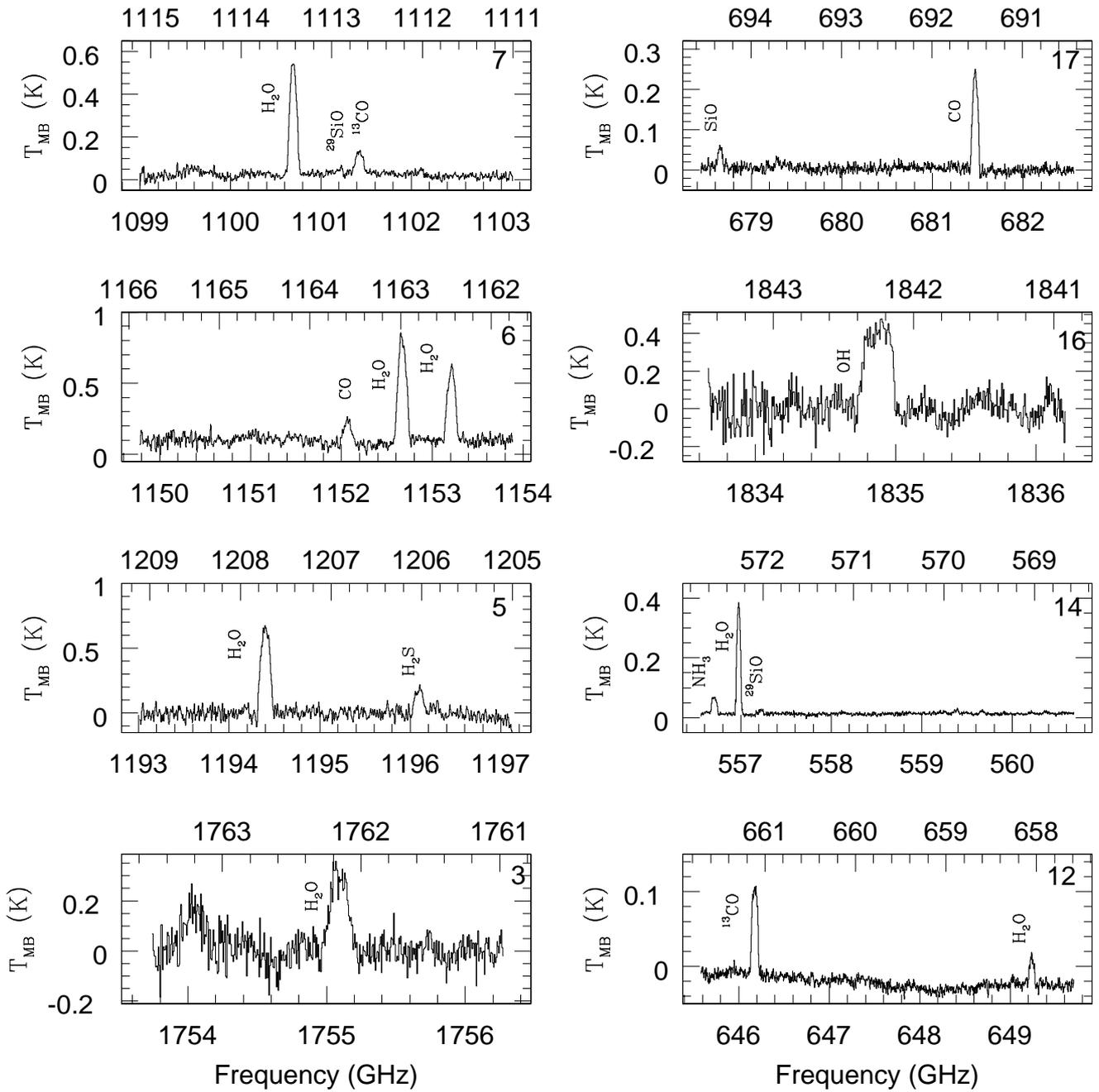}}
\caption{HIFI spectra of AFGL~5379.
}
\end{figure*}

\begin{figure*}
\resizebox{\hsize}{!}{\includegraphics[width=7cm]{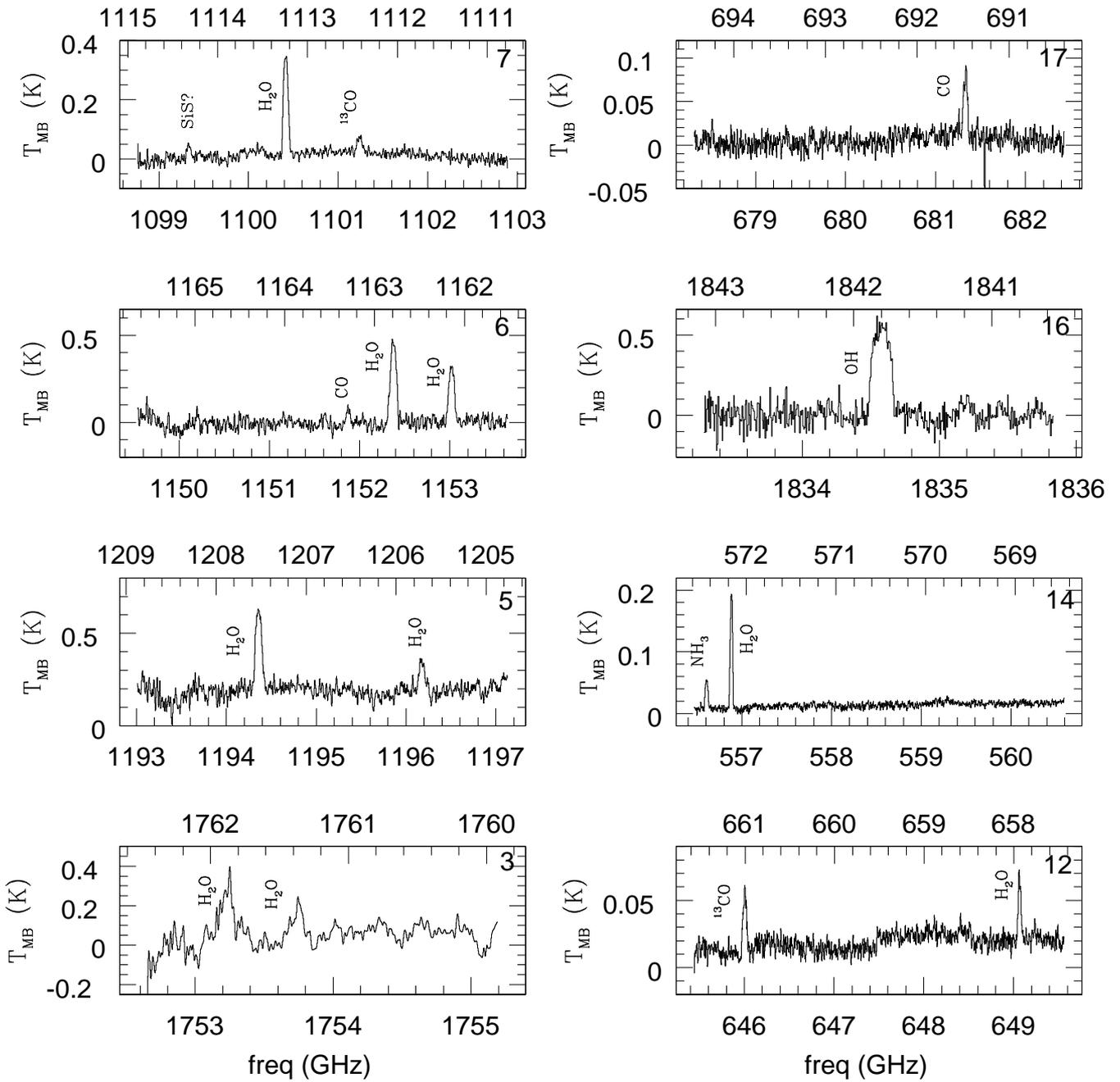}}
\caption{HIFI spectra of OH~26.5+0.6.
}
\end{figure*}

\begin{figure*}
\resizebox{\hsize}{!}{\includegraphics[width=7cm]{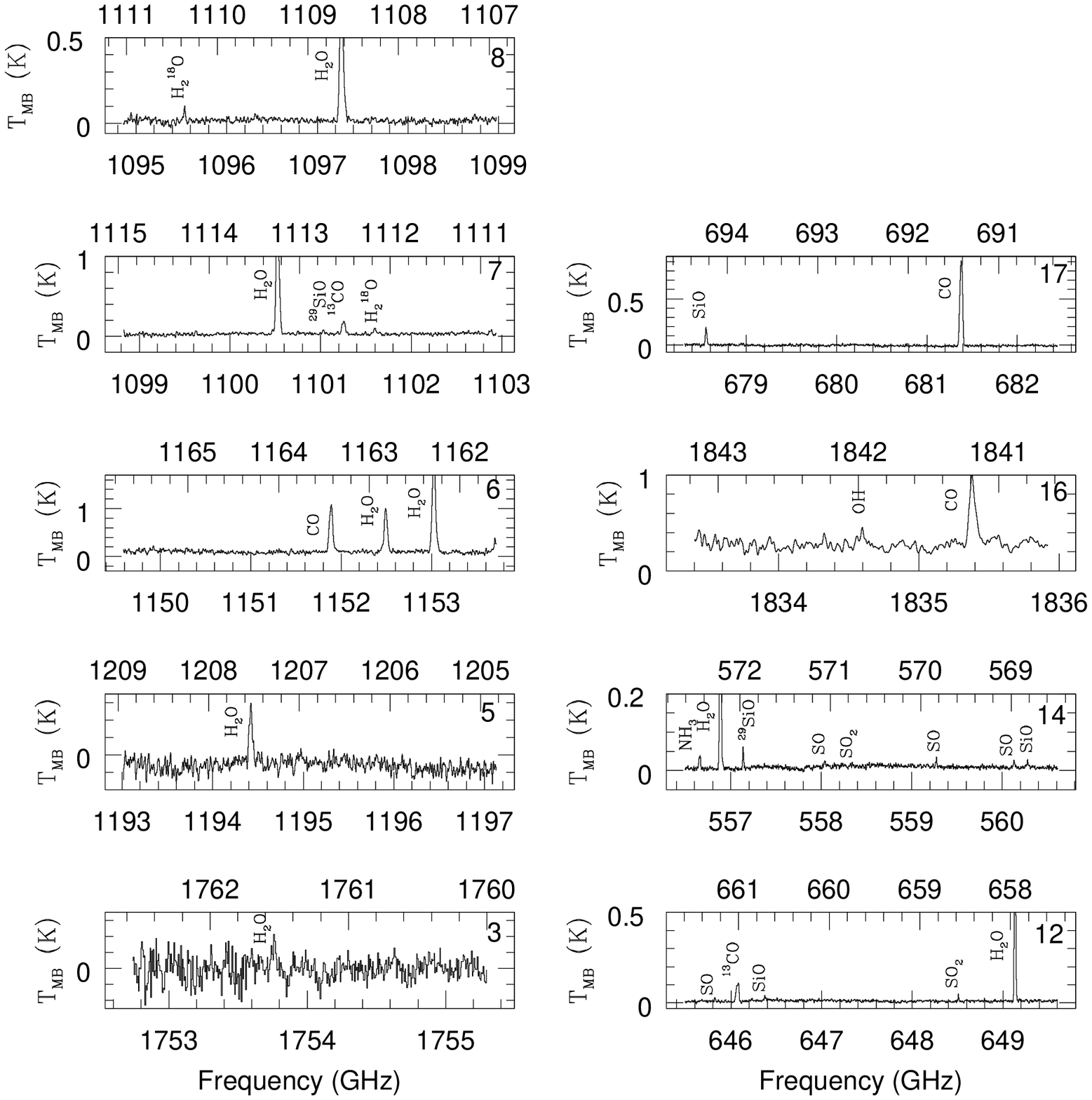}}
\caption{HIFI spectra of R Cas.
}
\label{fig_rcas}
\end{figure*}

\begin{acknowledgements}
HCSS / HSpot / HIPE is a joint development (are joint developments) 
by the Herschel Science Ground Segment Consortium, consisting of ESA, 
the NASA Herschel Science Center, and the HIFI, PACS and SPIRE consortia.
K.J., F.S, M.M., and H.O.\ acknowledge funding from the Swedish National
Space Board.
This work has been partially supported by the
Spanish MICINN, within the program CONSOLIDER INGENIO 2010, under grant
``Molecular Astrophysics: The Herschel and Alma Era -- ASTROMOL" (ref.:
CSD2009-00038). R.Sz.\ and M.Sch.\ acknowledge support from grant N 203
581040 from Polish MNiSW. J.C.\ thanks funding from MICINN, 
grant AYA2009-07304. This research was performed, in part, through a 
JPL contract funded by the National Aeronautics and Space Administration.

We would like to thank the referee for his/her careful reading 
and suggestions which improved the manuscript.
\end{acknowledgements}

\bibliographystyle{aa}
\bibliography{ref}

\onltab{1}{
\clearpage 
\onecolumn
\begin{landscape}
\begin{table*}
%\begin{longtable}{lcrccccccccc}
\caption{Calculated line intensities (K km s$^{-1}$) of the
observed lines. The setting numbers are internal HIFISTARS
LO frequency settings (see also Figs~\ref{fig_spec}-\ref{fig_rcas}).
The `-' marks the settings which have
not been observed and `x' indicates that the line is not detected
above the noise level.}
\label{lines_id}
\begin{tabular}{lcrccccccccc}
\hline \hline
  Molecule  & $\nu$ & $E_{\rm u}$ &
\multicolumn{9}{c}{I (K km s$^{-1}$)} \\
 & (GHz) & (K) & IRC+10011 & {\it o} Cet & IK Tau & R Dor & TX Cam & 
W Hya &AFGL~5379&OH~26.5& R Cas
\\
\hline % setting 1
setting 1 & & & & & & & & & & \\
o-H$_{2}$O  (5$_{3,2}$-5$_{2,3}$) & 1867.749 & 732.1 &
 7.9$\pm$1.5 &      -      & 8.9$\pm$2.2 &      -      &      -      &
 5.3$\pm$0.6 &      -      &      -      &      -      \\
\hline % setting 2
setting 2 & & & & & & & & & & \\
o-H$_{2}$O   (7$_{3,4}$-7$_{2,5}$) & 1797.159 & 1212.0 &
      -      &      -      &      -      &      -      &        -    &
 2.3$\pm$0.3 &      -      &      -      &      -     \\
\hline % setting 3
setting 3 & & & & & & & & & & \\
p-H$_{2}$O   (6$_{3,3}$-6$_{2,4}$) & 1762.043 & 951.8 &
 3.2$\pm$0.9 & 1.0$\pm$0.7 & 2.1$\pm$0.5 & 7.8$\pm$0.9 &      x      &
 2.4$\pm$0.6 & 8.9$\pm$1.8 & 5.5$\pm$1.4 & 0.4$\pm$0.3 \\
$^{13}$CO    (16-15)           & 1760.486 & 718.7 &
      x      & 1.3$\pm$0.5 &      x      & 2.8$\pm$0.7 &      x      &
      x      &      x      &      x      &      x      \\
o-H$_{2}$O  ($\nu_{2}$ 2$_{1,2}$-1$_{0,1}$) & 1753.914 & 2412.9 &
      x      & 1.4$\pm$0.5 & 1.9$\pm$0.4 & 7.0$\pm$1.0 &      x      &
 2.2$\pm$0.6 &      x      &      x      &      x      \\
\hline % setting 4
setting 4 & & & & & & & & & & \\
o-H$_{2}^{17}$O (3$_{0,3}$-2$_{1,2}$) & 1718.119 & 196.4 &
      x      &      -      &      x      &      -      &      -      &
 1.0$\pm$0.2 &      -      &      -      &      -      \\
p-H$_{2}$O   (5$_{3,3}$-6$_{0,6}$) & 1716.957 & 725.1 &
      x      &      -      & 2.7$\pm$0.7 &      -      &      -      &
      x      &      -      &      -      &      -      \\
o-H$_{2}$O   (3$_{0,3}$-2$_{1,2}$) & 1716.769 & 196.8 &
14.9$\pm$1.3 &       -     &26.0$\pm$1.0 &      -      &      -      &
24.9$\pm$0.3 &      -      &      -      &      -      \\
\hline % setting 5
setting 5 & & & & & & & & & & \\
p-H$_{2}$O   (4$_{2,2}$-4$_{1,3}$) & 1207.639 & 454.3 &
 7.6$\pm$2.0 & 1.2$\pm$0.3 & 8.9$\pm$0.8 & 7.0$\pm$0.5 &      x      &
 3.3$\pm$0.5 &22.6$\pm$2.0 & 9.2$\pm$1.1 & 3.1$\pm$0.4 \\
p-H$_{2}$O   ($\nu_{2}$ 1$_{1,1}$-0$_{0,0}$) & 1205.789 & 2352.4 &
      x      &      x      & 0.9$\pm$0.3 & 1.6$\pm$0.7 &      x      &
      x      &      x      &  3.8$\pm$0.9 &       x     \\
H$_{2}$S     (3$_{1,2}-2_{2,1}$) & 1196.012 & 136.8 &
      x      &      x      &      x      &      x      &      x      &
      x      & 4.4$\pm$1.6 &      x      &      x      \\
\hline % setting 6
setting 6 & & & & & & & & & & \\
o-H$_{2}$O   (3$_{2,1}$-3$_{1,2}$) & 1162.911 & 305.2 &
10.4$\pm$1.0 & 1.3$\pm$0.3 &13.9$\pm$0.8 &17.0$\pm$0.3 & 5.4$\pm$1.03&
 8.1$\pm$0.3 &22.8$\pm$0.7 &10.5$\pm$0.9 & 9.6$\pm$0.5 \\
o-H$_{2}$O   (3$_{1,2}$-2$_{2,1}$) & 1153.127 & 249.4 &
10.4$\pm$1.1 & 2.5$\pm$0.3 &26.1$\pm$1.0 &47.2$\pm$0.2 &10.3$\pm$0.8 &
21.1$\pm$0.4 &14.6$\pm$1.1 & 6.9$\pm$0.5 &22.9$\pm$0.4 \\
CO           (10-9)            & 1151.985 & 304.2 &
 5.2$\pm$1.1 &14.6$\pm$0.2 &10.7$\pm$0.8 &15.0$\pm$0.4 & 9.7$\pm$1.2 &
10.4$\pm$0.5 & 4.9$\pm$0.2 & 1.2$\pm$0.3 &11.8$\pm$0.2 \\
SO$_{2}$     (15$_{9,7} - 14_{8,6}$)   & 1151.852 & 308.6 & 
      x      &      x      &      x      & 1.1$\pm$0.4 &      x      &
      x      &      x      &      x      &              \\
HCN         (13-12)          & 1151.452 & 386.9 & 
      x      &      x      &      x      & 1.9$\pm$0.3 &      x      &
 1.4$\pm$0.3 &      x      &      x      &              \\
\hline % setting 7
setting 7 & & & & & & & & & & \\
SiS      (v=1 J=62-61)?        & 1114.431 & 2760.2 &
      x      &      x      &      x      &      x      &      x      &
      x      &      x      & 0.6$\pm$0.2 &      x      \\
SO$_{2}$ (13$_{9,5}$-12$_{8,4}$)  & 1113.506 &  195.9 &
      x      &      x      &      x      & 0.6$\pm$0.1 &      x      &
      x      &      x      &      x      &      x      \\
p-H$_{2}$O   (1$_{1,1}$-0$_{0,0}$) & 1113.343 & 53.4 &
 9.5$\pm$0.4 & 2.3$\pm$0.1 &24.4$\pm$0.5 &28.8$\pm$0.2 &11.3$\pm$0.7 &
15.4$\pm$0.1 &12.3$\pm$0.4 & 5.7$\pm$0.2 &17.0$\pm$0.1 \\
$^{29}$SiO   (26-25)           & 1112.833 & 721.6 &
      x      & 0.6$\pm$0.1 & 1.0$\pm$0.3 & 2.2$\pm$0.2 &      x      &
 0.9$\pm$0.2 & 0.5$\pm$0.2 &      x      & 0.3$\pm$0.1 \\
SO$_{2}$ (23$_{7,17}$-22$_{6,16}$) & 1102.115 &  371.6 &
      x      &      x      &      x      & 0.6$\pm$0.2&      x      &
      x      &      x      &      x      &      x      \\
p-H$_{2}^{18}$O (1$_{1,1}$-0$_{0,0}$) & 1101.698 & 52.9 &
 2.9$\pm$0.4 & 0.4$\pm$0.1 & 3.5$\pm$0.3 & 1.6$\pm$0.1 &      x      &
 1.5$\pm$0.1 &      x      &      x      & 0.7$\pm$0.1 \\
$^{13}$CO    (10-9)            & 1101.349 & 290.8 &
 1.2$\pm$0.7 & 2.8$\pm$0.1 & 2.3$\pm$0.4 & 2.5$\pm$0.1 &      x      &
 1.3$\pm$0.2 & 2.7$\pm$0.4 & 1.0$\pm$0.3 & 1.8$\pm$0.3 \\
$^{30}$SiO   (26-25)           & 1099.708 & 713.1 &
      x      & 0.7$\pm$0.1 &      x      & 1.9$\pm$0.3 &      x      &
 0.9$\pm$0.2 &      x      &      x      &      x      \\
\hline % setting 8
setting 8 & & & & & & & & & & \\
p-H$_{2}^{17}$O (1$_{1,1}$-0$_{0,0}$) & 1107.167 & 53.1 &
 2.0$\pm$0.3 &      x      & 1.1$\pm$0.3 & 0.8$\pm$0.2 &      x      &
 0.5$\pm$0.1 &      -      &      -      &      x      \\
o-H$_{2}$O   (3$_{1,2}$-3$_{0,3}$) & 1097.365 & 249.4 &
 7.2$\pm$0.4 & 1.8$\pm$0.2 &13.5$\pm$0.3 &15.8$\pm$0.1 & 6.5$\pm$0.4 &
 8.0$\pm$0.2 &      -      &      -      &11.2$\pm$0.2 \\
o-H$_{2}^{17}$O (3$_{1,3}$-3$_{0,3}$) & 1096.414 & 249.1 &
 1.2$\pm$0.4 & 0.2$\pm$0.2 & 1.0$\pm$0.2 & 1.0$\pm$0.2 &      x      &
 0.5$\pm$0.1 &      -      &      -      &      x      \\
o-H$_{2}^{18}$O (3$_{1,3}$-3$_{0,3}$) & 1095.627 & 248.7 &
 1.8$\pm$0.5 & 0.3$\pm$0.1 & 2.7$\pm$0.4 & 1.9$\pm$0.2 &      x      &
 1.5$\pm$0.1 &      -      &      -      & 0.5$\pm$0.1 \\
\hline % setting 9
setting 9 & & & & & & & & & & \\
%p-H$_{2}$O   & $\nu_{2} 2_{1,1}$-2$_{0,2}$ & 1000.850 & 3255.3453 & \\
$^{13}$CO    (9-8)            & 991.329  & 237.9 &
 1.9$\pm$0.4 &      -      & 2.4$\pm$0.5 &      -      &      -      &
 1.2$\pm$0.2 &      -      &      -      &      -      \\
SiO         (v=1; J=23-22)      & 990.355  & 2339.9 &
      x      &      -      &      x      &      -      &      -      &
 0.4$\pm$0.1 &      -      &      -      &      -     \\
SO          (23$_{24}-22_{23}$)   & 988.616  & 574.6 &
      x      &      -      &      x      &      -      &      -      &
 0.6$\pm$0.1 &      -      &      -      &      -      \\
p-H$_{2}$O   (2$_{0,1}$-1$_{1,1}$) & 987.927  & 100.8 &
10.7$\pm$0.6 &      -      &23.7$\pm$0.6 &      -      &      -      &
18.6$\pm$0.2 &      -      &      -      &      -      \\
\hline
} % end of onltab
\clearpage
\onltab{1}{
\end{tabular}
\end{table*}
\setcounter{table}{0}
\begin{table*}
\caption{Cont.}
%\tiny
\begin{tabular}{lcrccccccccc}
\hline \hline
  Molecule  & $\nu$ & $E_{\rm u}$ &
\multicolumn{9}{c}{I (K km s$^{-1}$)} \\
 & (GHz) & (K) & IRC+10011 & {\it o} Cet & IK Tau & R Dor & TX Cam & 
W Hya &AFGL~5379&OH~26.5& R Cas\\
\hline % setting 10
setting 10 & & & & & & & & & & \\
p-H$_{2}$O   (5$_{2,4}$-4$_{3,1}$) & 970.315  & 598.8 &
 7.0$\pm$0.4 &      -      &17.5$\pm$0.2 &      -      &      -      &
20.9$\pm$0.1 &      -      &      -      &      -      \\
\hline % setting 11
setting 11 & & & & & & & & & & \\
p-H$_{2}$O   (2$_{1,1}$-2$_{0,2}$) & 752.033  & 136.9 &
 6.7$\pm$0.4 &      -      &13.2$\pm$0.2 &      -      &      -      &
 7.0$\pm$0.1 &      -      &      -      &      -      \\
\hline % setting 12
setting 12 & & & & & & & & & & \\
$^{13}$CO    (6-5)             & 661.067  & 111.0 &
 1.1$\pm$0.2 & 3.0$\pm$0.1 & 2.6$\pm$0.3 & 2.7$\pm$0.1 & 1.4$\pm$0.2 &
 1.1$\pm$0.1 & 3.8$\pm$0.2 & 1.1$\pm$0.2 & 1.5$\pm$0.2 \\
SO$_{2}$     (24$_{7,17}-24_{6,18}$) & 659.898 & 396.0 &
      x      &      x      &      x      & 0.2$\pm$0.1 &      x      &
      x      &      x      &      x      &      x      \\
SO$_{2}$     (37$_{1,37}-36_{0,36}$) & 659.421 & 608.8 &
      x      & 0.1$\pm$0.05 &      x      & 0.5$\pm$0.1 &     x      &
 0.1$\pm$0.03 &       x      &      x      &      x      \\
SO$_{2}$     (36$_{1,35}-35_{2,34}$) & 658.632 & 605.4 &
      x      &      x      & 0.1$\pm$0.04 & 0.5$\pm$0.1 &     x      &
 0.4$\pm$0.1 &       x     &      x       & 0.2$\pm$0.03 \\
C$^{18}$O    (6-5)                 & 658.553 & 111 &
      x      &      x      &      x      & 0.3$\pm$0.1 &      x      &
      x      &      x      &      x      &      x      \\
o-H$_{2}$O   ($\nu_{2} 1_{1,0}$-1$_{0,1}$) & 658.007 & 2360.3 &
 2.9$\pm$0.1 & 5.3$\pm$0.1 & 9.2$\pm$0.1 &42.7$\pm$0.2 & 2.0$\pm$0.1 &
28.9$\pm$0.1 & 1.1$\pm$0.2 & 0.8$\pm$0.1 &16.9$\pm$0.1 \\
SiO         (v=1; J=15-14)      & 646.429 & 2017.4 &
      x      &      x       &      x      & 0.5$\pm$0.1 &      x      &
 0.2$\pm$0.04&       x      &      x      & 0.1$\pm$0.04\\
SO           (15$_{16}-14_{15}$)  & 645.875 & 252.6 &
      x      &      x       &      x      & 1.2$\pm$0.1 &      x      &
 0.4$\pm$0.1 &       x      &      x      & 0.1$\pm$0.03 \\
\hline % setting 13
setting 13 & & & & & & & & & & \\
o-H$_{2}$O   (5$_{3,2}$-4$_{4,1}$) & 620.701 & 732.1 &
 1.7$\pm$0.1 &       -      &15.0$\pm$0.1 &      -      &      -      &
 8.4$\pm$0.1 &       -      &       -     &      -      \\ 
HCN         (7-6)             & 620.304 & 119.1 &
 1.0$\pm$0.1 &       -      & 2.6$\pm$0.1 &      -      &      -      &
 2.2$\pm$0.1 &       -      &       -     &      -      \\
SiO         (14-13)           & 607.599 & 218.8 &
 0.7$\pm$0.3 &       -      & 2.2$\pm$0.1 &      -      &      -      &
 2.8$\pm$0.1 &       -      &       -     &      -      \\
%\hline 
%\hline
\hline% setting 14
setting 14 & & & & & & & & & & \\
NH$_{3}$     (1$_{0}$-0$_{0}$)      & 572.498 & 27.5 &
 2.2$\pm$0.2 &       x     & 4.7$\pm$0.1 & 0.5$\pm$0.1 & 1.5$\pm$0.1 &
 0.5$\pm$0.1 & 1.7$\pm$0.1 & 0.8$\pm$0.1 & 0.3$\pm$0.1 \\
SO$_{2}$     (32$_{0,32}-31_{1,31}$) & 571.553 & 459.0 &
       x     &       x     &       x     & 0.3$\pm$0.1 &       x     &
       x     &       x     &       x     &       x     \\
SO$_{2}$     (32$_{2,30}-31_{3,29}$) & 571.532 & 504.3 &
       x     &       x     &       x     & 0.1$\pm$0.03 &       x     &
       x     &       x     &       x     &       x     \\
SO$_{2}$     (16$_{6,10}-16_{5,11}$) & 560.613 & 213.3 &
       x     &       x     &       x     & 0.2$\pm$0.04 &       x     &
 0.1$\pm$0.02 &       x     &       x     &       x     \\
SiO         (v=1; J=13-12)        & 560.326 & 1957.4 &
       x     & 0.1$\pm$0.02& 0.4$\pm$0.1 & 0.5$\pm$0.1 &       x      &
 0.3$\pm$0.1 & 0.2$\pm$0.1 &       x     & 0.1$\pm$0.02 \\
SO          ($13_{14}-12_{13}$)     & 560.178 &  192.7 &
       x     & 0.1$\pm$0.02& 0.1$\pm$0.03& 1.2$\pm$0.1 &       x     &
 0.5$\pm$0.1 &       x     &       x     & 0.1$\pm$0.03 \\
SO$_{2}$     (18$_{6,12}-18_{5,13}$) & 559.882 & 245.5 &
       x     &       x     &       x     & 0.1$\pm$0.04&       x     &
 0.1$\pm$0.03&       x     &       x     &       x      \\
SO          (13$_{13}-12_{12}$)     & 559.319  & 201.1 &
       x     & 0.1$\pm$0.02& 0.1$\pm$0.03& 1.1$\pm$0.1 &       x     &
 0.4$\pm$0.1 & 0.3$\pm$0.1 &       x     & 0.1$\pm$0.03 \\
SO$_{2}$     (20$_{6,14}-20_{5,15}$) & 558.812 & 281.4 &
       x     &       x     &       x     & 0.1$\pm$0.02&       x     &
 0.1$\pm$0.04&       x     &       x     &       x      \\
SO$_{2}$     (21$_{6,16}-21_{5,17}$) & 558.391 & 300.8 &
       x     &       x     & 0.1$\pm$0.07& 0.2$\pm$0.03&       x     &
 0.2$\pm$0.03&       x     &       x     &0.04$\pm$0.02 \\
SO          (13$_{12}-12_{11}$)     & 558.087  & 194.4 &
       x     & 0.1$\pm$0.03& 0.1$\pm$0.1 & 1.1$\pm$0.1 &       x     &
 0.3$\pm$0.04&       x     &       x     & 0.1$\pm$0.04\\
SO$_{2}$     (22$_{6,16}-22_{5,17}$) & 557.283 & 321.0 &
       x     &       x     &       x     & 0.2$\pm$0.1 &       x     &
 0.1$\pm$0.06&       x     &       x     &       x     \\
$^{29}$SiO   (13-12)             & 557.179  & 187.2 &
 0.2$\pm$0.1 & 0.3$\pm$0.1 & 0.6$\pm$0.1 & 2.0$\pm$0.1 & 0.2$\pm$0.1 &
 1.0$\pm$0.1 & 0.5$\pm$0.2 &       x     & 0.4$\pm$0.04\\
o-H$_{2}$O   (1$_{1,0}$-1$_{0,1}$)   & 556.936 & 61.0 &
 7.1$\pm$0.1 & 1.1$\pm$0.1 & 9.4$\pm$0.1 & 10.2$\pm$0.1& 5.8$\pm$0.1 &
 5.4$\pm$0.1 & 9.7$\pm$0.1 & 3.1$\pm$0.1 &  6.9$\pm$0.1 \\
\hline % setting 16
setting 16 & & & & & & & & & & \\
CO          (16-15)             & 1841.345& 751.7 &
 3.9$\pm$0.9 &  7.0$\pm$0.2 &12.9$\pm$1.8 & 9.1$\pm$0.5& 0.6$\pm$0.2 & 
 5.8$\pm$0.3 &       x     &       x     & 5.5$\pm$0.4 \\
%OH          &(2,1,-1,0,1)-(1,1,1,0)& 1834.735 & 126.2935 & \\
OH          ($^{2}\pi_{1/2}$ J=3/2-1/2) & 1834.747 & 181.7 &
 3.4$\pm$2.1 & 1.5$\pm$0.3 & 4.2$\pm$1.5 & 3.1$\pm$0.5 &       x     &
 2.2$\pm$0.4 &11.6$\pm$2.1 &11.2$\pm$1.4 & 2.0$\pm$0.7 \\
%OH          &(2,1,-1,0,1)-(1,1,1,0)& 1834.750 & 126.2930 & \\
\hline % setting 17
setting 17 & & & & & & & & & & \\
SiO         (16-15)             & 694.275   & 283.3 &
 0.5$\pm$0.3 & 0.5$\pm$0.1 & 2.1$\pm$0.3 & 5.7$\pm$0.1 & 0.9$\pm$0.2 &
 2.9$\pm$0.1 & 1.3$\pm$0.4 &       x     & 1.7$\pm$0.1 \\
CO          (6-5)               & 691.473   & 116.2 &
 4.4$\pm$0.4 &16.2$\pm$0.1 &11.6$\pm$0.4 &14.9$\pm$0.2 &14.2$\pm$0.3 &
 9.6$\pm$0.1 & 7.8$\pm$0.4 & 1.7$\pm$0.3 &14.7$\pm$0.2 \\
H$^{13}$CN   (8-7)               & 690.551   & 149.1 &
       x     &       x     &       x     &       x     &       x     &
 0.3$\pm$0.1 &       x     &       x     &       x     \\
\hline
\end{tabular}
\end{table*}
%\end{longtable}
\end{landscape}
} % end of onltab

\end{document}